\documentclass[aps,prl,twocolumn,superscriptaddress,showpacs,preprintnumbers,amsmath,amssymb]{revtex4-1}
\usepackage[utf8]{inputenc}
\usepackage{graphicx}
\usepackage{dcolumn}
\usepackage{bm}
\usepackage{notes2bib}
\usepackage{graphicx}
\usepackage{gensymb}
\usepackage{url}
\usepackage{amsmath}
\usepackage{latexsym}
\usepackage{amssymb}
\usepackage[dvipsnames]{xcolor}
\usepackage{slashed}
\usepackage[normalem]{ulem}
\usepackage[citecolor=blue]{hyperref}
\usepackage{slantsc}

\begin{document}


\title{Measurement of branching fractions and $CP$ asymmetries for $D_s^{+} \rightarrow K^+ (\eta, \pi^0) $ and $D_s^{+} \rightarrow \pi^+ (\eta, \pi^0)$ decays at Belle}

\affiliation{Department of Physics, University of the Basque Country UPV/EHU, 48080 Bilbao}
\affiliation{University of Bonn, 53115 Bonn}
\affiliation{Brookhaven National Laboratory, Upton, New York 11973}
\affiliation{Budker Institute of Nuclear Physics SB RAS, Novosibirsk 630090}
\affiliation{Faculty of Mathematics and Physics, Charles University, 121 16 Prague}
\affiliation{Chonnam National University, Gwangju 61186}
\affiliation{University of Cincinnati, Cincinnati, Ohio 45221}
\affiliation{Deutsches Elektronen--Synchrotron, 22607 Hamburg}
\affiliation{Duke University, Durham, North Carolina 27708}
\affiliation{Department of Physics, Fu Jen Catholic University, Taipei 24205}
\affiliation{Key Laboratory of Nuclear Physics and Ion-beam Application (MOE) and Institute of Modern Physics, Fudan University, Shanghai 200443}
\affiliation{Justus-Liebig-Universit\"at Gie\ss{}en, 35392 Gie\ss{}en}
\affiliation{Gifu University, Gifu 501-1193}
\affiliation{II. Physikalisches Institut, Georg-August-Universit\"at G\"ottingen, 37073 G\"ottingen}
\affiliation{SOKENDAI (The Graduate University for Advanced Studies), Hayama 240-0193}
\affiliation{Gyeongsang National University, Jinju 52828}
\affiliation{Department of Physics and Institute of Natural Sciences, Hanyang University, Seoul 04763}
\affiliation{University of Hawaii, Honolulu, Hawaii 96822}
\affiliation{High Energy Accelerator Research Organization (KEK), Tsukuba 305-0801}
\affiliation{J-PARC Branch, KEK Theory Center, High Energy Accelerator Research Organization (KEK), Tsukuba 305-0801}
\affiliation{Higher School of Economics (HSE), Moscow 101000}
\affiliation{Forschungszentrum J\"{u}lich, 52425 J\"{u}lich}
\affiliation{IKERBASQUE, Basque Foundation for Science, 48013 Bilbao}
\affiliation{Indian Institute of Science Education and Research Mohali, SAS Nagar, 140306}
\affiliation{Indian Institute of Technology Guwahati, Assam 781039}
\affiliation{Indian Institute of Technology Hyderabad, Telangana 502285}
\affiliation{Indian Institute of Technology Madras, Chennai 600036}
\affiliation{Indiana University, Bloomington, Indiana 47408}
\affiliation{Institute of High Energy Physics, Chinese Academy of Sciences, Beijing 100049}
\affiliation{Institute of High Energy Physics, Vienna 1050}
\affiliation{Institute for High Energy Physics, Protvino 142281}
\affiliation{INFN - Sezione di Napoli, 80126 Napoli}
\affiliation{INFN - Sezione di Torino, 10125 Torino}
\affiliation{Advanced Science Research Center, Japan Atomic Energy Agency, Naka 319-1195}
\affiliation{J. Stefan Institute, 1000 Ljubljana}
\affiliation{Institut f\"ur Experimentelle Teilchenphysik, Karlsruher Institut f\"ur Technologie, 76131 Karlsruhe}
\affiliation{Kavli Institute for the Physics and Mathematics of the Universe (WPI), University of Tokyo, Kashiwa 277-8583}
\affiliation{Department of Physics, Faculty of Science, King Abdulaziz University, Jeddah 21589}
\affiliation{Kitasato University, Sagamihara 252-0373}
\affiliation{Korea Institute of Science and Technology Information, Daejeon 34141}
\affiliation{Korea University, Seoul 02841}
\affiliation{Kyungpook National University, Daegu 41566}
\affiliation{Universit\'{e} Paris-Saclay, CNRS/IN2P3, IJCLab, 91405 Orsay}
\affiliation{P.N. Lebedev Physical Institute of the Russian Academy of Sciences, Moscow 119991}
\affiliation{Faculty of Mathematics and Physics, University of Ljubljana, 1000 Ljubljana}
\affiliation{Ludwig Maximilians University, 80539 Munich}
\affiliation{Luther College, Decorah, Iowa 52101}
\affiliation{Malaviya National Institute of Technology Jaipur, Jaipur 302017}
\affiliation{University of Maribor, 2000 Maribor}
\affiliation{Max-Planck-Institut f\"ur Physik, 80805 M\"unchen}
\affiliation{School of Physics, University of Melbourne, Victoria 3010}
\affiliation{University of Mississippi, University, Mississippi 38677}
\affiliation{Moscow Physical Engineering Institute, Moscow 115409}
\affiliation{Graduate School of Science, Nagoya University, Nagoya 464-8602}
\affiliation{Kobayashi-Maskawa Institute, Nagoya University, Nagoya 464-8602}
\affiliation{Universit\`{a} di Napoli Federico II, 80126 Napoli}
\affiliation{Nara Women's University, Nara 630-8506}
\affiliation{National Central University, Chung-li 32054}
\affiliation{National United University, Miao Li 36003}
\affiliation{Department of Physics, National Taiwan University, Taipei 10617}
\affiliation{H. Niewodniczanski Institute of Nuclear Physics, Krakow 31-342}
\affiliation{Nippon Dental University, Niigata 951-8580}
\affiliation{Niigata University, Niigata 950-2181}
\affiliation{University of Nova Gorica, 5000 Nova Gorica}
\affiliation{Novosibirsk State University, Novosibirsk 630090}
\affiliation{Okinawa Institute of Science and Technology, Okinawa 904-0495}
\affiliation{Osaka City University, Osaka 558-8585}
\affiliation{Pacific Northwest National Laboratory, Richland, Washington 99352}
\affiliation{Panjab University, Chandigarh 160014}
\affiliation{Peking University, Beijing 100871}
\affiliation{University of Pittsburgh, Pittsburgh, Pennsylvania 15260}
\affiliation{Research Center for Nuclear Physics, Osaka University, Osaka 567-0047}
\affiliation{Meson Science Laboratory, Cluster for Pioneering Research, RIKEN, Saitama 351-0198}
\affiliation{Department of Modern Physics and State Key Laboratory of Particle Detection and Electronics, University of Science and Technology of China, Hefei 230026}
\affiliation{Seoul National University, Seoul 08826}
\affiliation{Showa Pharmaceutical University, Tokyo 194-8543}
\affiliation{Soochow University, Suzhou 215006}
\affiliation{Soongsil University, Seoul 06978}
\affiliation{Sungkyunkwan University, Suwon 16419}
\affiliation{School of Physics, University of Sydney, New South Wales 2006}
\affiliation{Department of Physics, Faculty of Science, University of Tabuk, Tabuk 71451}
\affiliation{Tata Institute of Fundamental Research, Mumbai 400005}
\affiliation{Department of Physics, Technische Universit\"at M\"unchen, 85748 Garching}
\affiliation{School of Physics and Astronomy, Tel Aviv University, Tel Aviv 69978}
\affiliation{Toho University, Funabashi 274-8510}
\affiliation{Department of Physics, Tohoku University, Sendai 980-8578}
\affiliation{Earthquake Research Institute, University of Tokyo, Tokyo 113-0032}
\affiliation{Department of Physics, University of Tokyo, Tokyo 113-0033}
\affiliation{Virginia Polytechnic Institute and State University, Blacksburg, Virginia 24061}
\affiliation{Wayne State University, Detroit, Michigan 48202}
\affiliation{Yamagata University, Yamagata 990-8560}
\affiliation{Yonsei University, Seoul 03722}
  \author{Y.~Guan}\affiliation{University of Cincinnati, Cincinnati, Ohio 45221} 
  \author{A.~J.~Schwartz}\affiliation{University of Cincinnati, Cincinnati, Ohio 45221} 
  \author{K.~Kinoshita}\affiliation{University of Cincinnati, Cincinnati, Ohio 45221} 
  \author{I.~Adachi}\affiliation{High Energy Accelerator Research Organization (KEK), Tsukuba 305-0801}\affiliation{SOKENDAI (The Graduate University for Advanced Studies), Hayama 240-0193} 
  \author{H.~Aihara}\affiliation{Department of Physics, University of Tokyo, Tokyo 113-0033} 
  \author{S.~Al~Said}\affiliation{Department of Physics, Faculty of Science, University of Tabuk, Tabuk 71451}\affiliation{Department of Physics, Faculty of Science, King Abdulaziz University, Jeddah 21589} 
  \author{D.~M.~Asner}\affiliation{Brookhaven National Laboratory, Upton, New York 11973} 
  \author{H.~Atmacan}\affiliation{University of Cincinnati, Cincinnati, Ohio 45221} 
  \author{V.~Aulchenko}\affiliation{Budker Institute of Nuclear Physics SB RAS, Novosibirsk 630090}\affiliation{Novosibirsk State University, Novosibirsk 630090} 
  \author{T.~Aushev}\affiliation{Higher School of Economics (HSE), Moscow 101000} 
% ...
\collaboration{The Belle Collaboration}

\begin{abstract}
We report measurements of the branching fractions and $CP$ asymmetries for $D_s^{+} \rightarrow K^{+}  \eta $, $D_s^{+} \rightarrow K^{+} \pi^0 $, and $D_s^{+} \rightarrow \pi^{+}  \eta $ decays, and the branching fraction for $D_s^{+} \rightarrow \pi^{+} \pi^0$.
Our results are based on a data sample 
corresponding to an integrated luminosity 
of 921 fb$^{-1}$ collected by the Belle detector at the 
KEKB $e^+e^-$ asymmetric-energy collider. Our measurements of 
$CP$ asymmetries in these decays are the most precise to-date; 
no evidence for $CP$ violation is found. 
\end{abstract}
\maketitle

Charm hadrons provide a unique opportunity to study charge-parity ($CP$) violation  in the up-quark sector.
Within the Standard Model (SM), $CP$ violation ({$CPV$}) in charm decays is expected to be small, at the level of~$10^{-3}$~\cite{Cheng:2012wr,Li}.
The largest effect is expected to occur in singly Cabibbo-suppressed (SCS) decays~\cite{Grossman:2006jg,Grossman:2012eb,Cheng:2012xb,Lenz:2020awd}, 
which receive a contribution from a ``penguin" (internal loop) diagram.
The only evidence for
$CPV$ in the charm sector thus far was obtained by the LHCb experiment~\cite{Aaij:2019kcg}, which measured SCS $D^0 \rightarrow K^+ K^-$ and $D^0 \rightarrow \pi^+ \pi^-$ decays. The LHCb result has generated
much interest in the literature~\cite{Dery:2019ysp,Cheng:2019ggx,Buccella:2019kpn}.
One can also search for $CPV$ in Cabibbo-favored (CF) decays; as these decays proceed via tree-level decay amplitudes, an observation of $CPV$ would be a clear sign of new physics.

Here we present improved measurements of the branching 
fractions and $CP$ asymmetries for charm 
decays~\footnote{Charge-conjugate modes are implicitly included throughout this paper unless stated otherwise.}
 $D_s^+ \rightarrow K^+ \eta$, $D_s^+ \rightarrow K^+ \pi^0$, $D_s^+ \rightarrow \pi^+ \eta$, and $D_s^+ \rightarrow \pi^+ \pi^0$. 
The first two modes are SCS decays, while $D_s^+ \rightarrow \pi^+ \eta$ is CF, and $D_s^+ \rightarrow \pi^+ \pi^0$ proceeds via an annihilation amplitude. For this last mode, the branching fraction is expected to be very small~\cite{Li}, and only an upper limit has been obtained from experiments for its value~\cite{CLEO}.
The most recent measurements of these branching fractions were made by the CLEO~\cite{CLEO} and BESIII~\cite{BESIII} experiments. 
Higher precision measurements would help improve theoretical predictions for $CPV$~\cite{Muller:2015rna,Li,Cheng:2012wr}.
The only
measurements of $CPV$ in these decays were made by the CLEO experiment~\cite{CLEO}; our measurements presented here have significantly improved precision.

We define the $CP$ asymmetry in the decay rates as
 \begin{equation}
A_{CP} = \frac{\Gamma(D_s^+ \rightarrow f) - \Gamma(D_s^-\rightarrow \bar{f})}{\Gamma(D_s^+ \rightarrow f) + \Gamma(D_s^-\rightarrow \bar{f})}\,,
\end{equation} 
where $\Gamma(D_s^+ \rightarrow f)$ and $\Gamma(D_s^- \rightarrow \bar{f})$ are the partial decay widths for the final state $f$ and its $CP$-conjugate state~$\bar{f}$.
As our measured $A_{CP}$ corresponds to charged $D$ mesons, which do not undergo mixing, a nonzero value would indicate {\it direct\/}~$CP$ violation~\cite{BigiSanda_book}.

Our measurements are based on data recorded by the Belle detector~\cite{belle_detector} running at the KEKB~\cite{KEKB} asymmetric-energy $e^+e^-$ collider. The data samples were collected 
at $e^+e^-$ center-of-mass (CM) energies corresponding to the $\Upsilon(4S)$ and $\Upsilon(5S)$ resonances, and at 60~MeV below the $\Upsilon(4S)$ resonance. The corresponding integrated luminosities are 711~fb$^{-1}$, 121~fb$^{-1}$, and 89~fb$^{-1}$, respectively.
The Belle detector is a large-solid-angle magnetic spectrometer consisting of a silicon vertex detector (SVD), a central drift chamber (CDC), an array of aerogel threshold Cherenkov counters (ACC), a barrel-like arrangement of time-of-flight scintillation counters (TOF), and an electromagnetic calorimeter (ECL) consisting of CsI(Tl) crystals. These components are all located inside a superconducting solenoid coil that provides a 1.5~T magnetic field. An iron flux-return located outside of the coil is instrumented to detect $K^0_L$ mesons and to identify muons.

We calculate signal reconstruction efficiencies, optimize selection criteria, and study various backgrounds using Monte Carlo (MC) simulated events. MC events are generated using  {{\sc{evtgen}}}~\cite{Lange:2001uf} and {{\sc{pythia}}}~\cite{Sjostrand:2000wi}, 
and they are subsequently processed through a detector simulation using {{\sc{geant3}}}~\cite{Brun:1987ma}. Final-state radiation from charged particles is implemented during event generation using the {{\sc{photos}}} package~\cite{PHOTOS}. 

Signal $D_s^{+}$ mesons are produced via the process $e^+e^-\rightarrow c \bar{c}$, where one of the two charm quarks hadronizes into a $D_s^{+}$ meson. 
We also search for a low-momentum photon
to reconstruct 
$D_s^{*+} \rightarrow D_s^{+} \gamma$ decays. 
Such events, in which a $D_s^{*+} \rightarrow D_s^{+} \gamma$ decay
is reconstructed, are referred to as the ``tagged" sample. Otherwise, in the case of no reconstructed $D_s^{*+}$ decay, events are referred to as the ``untagged" sample~\cite{Babu:2017bjn}.
The former has low backgrounds, while the latter has higher statistics. The tagged and untagged samples are statistically independent, i.e., a reconstructed $D_s^+$ candidate will be in one or the other but not in both.
Because the total number of $D_s^{+}$ produced in data is not 
precisely known, we measure the branching fractions of signal 
modes relative to that of the CF mode 
$D_s^{+} \rightarrow \phi\,(\rightarrow K^+K^-)\,\pi^+$,
which has high statistics.

Charged-track candidates are required to originate near 
the $e^+ e^-$ interaction point (IP) and have an impact 
parameter along the $z$ axis (defined as opposite the $e^+$ beam 
direction) of less than 4.0~cm, and in the $x$-$y$ (transverse)
plane of less than 1.0~cm. The
tracks are required to have a transverse momentum greater 
than 100 MeV/$c$. To identify pion and kaon candidates,
a particle identification likelihood is constructed based 
on energy-loss measurements in the CDC, time-of-flight information 
from the TOF, and  light yield measurements from the ACC~\cite{bellePID}.
A track is identified as a kaon if the ratio
$\mathcal{L}(K)/(\mathcal{L}(K) + \mathcal{L}(\pi))>0.6$,
where $\mathcal{L}(K)$ and $\mathcal{L}(\pi)$ are the likelihoods
that the track is a kaon or pion, respectively. If this criterion 
is not satisfied, the track is assumed to be a pion. 
The corresponding efficiencies are approximately 84\% for kaons and 94\% for pions.
Photon candidates are reconstructed from electromagnetic clusters in 
the ECL that do not have an associated charged track. Such candidates are
required to have an energy greater than 50~MeV in the barrel region, 
and greater than 100~MeV in the end-cap region. The hit times of energy 
deposited in the ECL must be consistent with the beam collision time, as calculated at the trigger level.
The photon energy deposited in the $3\times 3$ array 
of ECL crystals centered on the crystal with the highest energy is required to exceed 80\% of the energy deposited in the 
corresponding $5\times 5$ array of crystals.

Candidate $\pi^0$'s are reconstructed from photon pairs 
having an invariant mass satisfying
$0.120~{\rm GeV}/c^2 < M_{\gamma\gamma} < 0.150~{\rm GeV}/c^2$; 
this range corresponds to about $2.5\sigma$ in mass resolution.
Candidate $\eta$ mesons are reconstructed via $\eta \rightarrow \gamma \gamma$ ($\eta_{\gamma\gamma}$) and $\eta \rightarrow \pi^+ \pi^- \pi^0$ ($\eta_{3\pi}$) decays. 
To reduce combinatorial background from low-energy photons, $\pi^0$ and $\eta_{\gamma\gamma}$ candidates are required to have
$|E_{\gamma_1} - E_{\gamma_2}|/(E_{\gamma_1} + E_{\gamma_2})< 0.9$, where $E_{\gamma_1}$ and $E_{\gamma_2}$ are the energies of the two photons. If a photon can pair with another photon to form a $\pi^0$
candidate, then it is not used to reconstruct $\eta_{\gamma\gamma}$ candidates. The invariant masses of $\eta_{\gamma\gamma}$ and $\eta_{3\pi}$ candidates are required to satisfy 
$0.500~{\rm GeV}/c^2 < M_{\gamma\gamma} < 0.580~{\rm GeV}/c^2$ and 
$0.538~{\rm GeV}/c^2 < M_{\pi^+\pi^-\pi^0} < 0.557~{\rm GeV}/c^2$, respectively; 
these ranges correspond to 
about $3.0\sigma$ in mass resolution.
Mass-constrained fits are performed for $\pi^0$, $\eta_{\gamma\gamma}$, and $\eta_{3\pi}$ candidates to improve their momentum resolution.
For the reference mode $D_s^{+} \rightarrow   \phi \pi^{+} $, $\phi$
candidates are reconstructed from $K^+K^-$ pairs that form a 
vertex and have an invariant mass satisfying
$1.010~{\rm GeV}/c^2 < M_{K^+ K^-} < 1.030~{\rm GeV}/c^2$.
We also reconstruct $K^0_S\rightarrow\pi^+\pi^-$ decays, as 
the multiplicity of such decays (and also $K^+$ candidates) 
is used later by a neural network to reduce backgrounds.
These candidates are reconstructed from $\pi^+\pi^-$ pairs 
that form a vertex and satisfy 
$|M_{\pi^+\pi^-} - m^{}_{K^0_S}| < 20 ~{\rm MeV}/c^2$, 
where $m^{}_{K^0_S}$ is the nominal 
mass of the~${K^0_S}$~\cite{PDG2020}.

We subsequently reconstruct $D_s^+$ candidates by combining a 
$K^+$ or $\pi^+$ track with a $\pi^0$, $\eta_{\gamma\gamma}$, 
or $\eta_{3\pi}$ candidate. For $D_s^{+}\rightarrow \phi\pi^{+}$ 
decays, we combine a $\pi^+$ track with a $\phi$ candidate. 
For $D_s^+ \rightarrow (K^+,\pi^+) \pi^0$ and $D_s^+ \rightarrow \pi^+ \eta_{\gamma\gamma}$ decays, we require that the invariant mass satisfy
$1.86~{\rm GeV}/c^2 < M_{D_s^+} < 2.07~{\rm GeV}/c^2$;
for $D_s^+ \rightarrow K^+ (\eta_{\gamma\gamma}, \eta_{3\pi})$ and 
$D_s^+ \rightarrow \pi^+ \eta_{3\pi}$, we require 
$1.86~{\rm GeV}/c^2 < M_{D_s^+} < 2.05~{\rm GeV}/c^2$.
A narrower range is chosen for 
$D_s^+ \rightarrow K^+ (\eta_{\gamma\gamma}, \eta_{3\pi})$
in order to avoid an excess of 
events in the region $M>2.05$~GeV/$c^2$ originating 
from $D_s^+ \rightarrow \pi^+ \eta$ decays, with
the $\pi^{+}$ misidentified as a~$K^{+}$.
A narrower range is chosen for $D_s^+ \rightarrow \pi^+ \eta_{3\pi}$
due to its better resolution.

For the reference mode $D_s^+ \rightarrow \phi \pi^+$, 
we require $1.93~{\rm GeV}/c^2 < M_{D_s^+} < 2.01~{\rm GeV}/c^2$.
In addition, for $D_s^{+}\rightarrow (K^+,\pi^+)\eta_{3\pi}$ and 
$D_s^{+}\rightarrow \phi\pi^{+}$ decays, we require that
the charged tracks form a vertex.
To suppress combinatorial backgrounds and also $D_s^+$ candidates
originating from $B$ decays, we require that the $D_s^{+}$ momentum 
in the CM frame be greater than 2.3~GeV/$c$. 

We reconstruct $D_s^{*+}$ candidates by combining a $D_s^{+}$ 
candidate with a~$\gamma$. The $\gamma$ is required to have
an energy $E_{\gamma} > 0.15$~GeV and not be associated 
with a $\pi^0$ candidate. The mass difference 
$\Delta M \equiv M_{D_s^+\gamma} - M_{D_s^{+}}$, where $M_{D_s^+}$
is the invariant mass of the $D_s^+$ candidate, is required to
satisfy $0.125~{\rm GeV}/c^2 < \Delta M < 0.155~{\rm GeV}/c^2$.
The upper and lower ranges correspond
to about $2.5\sigma$ and $3.5\sigma$ in resolution, respectively. 
The lower range is larger due to a longer tail in the distribution 
for measurement of $\gamma$ energy.
The $D_s^{+}$ candidates that satisfy 
the above $D_s^{*+}\rightarrow D_s^+\gamma$ requirements
constitute the tagged sample.

To suppress backgrounds, 
we use a neural network~(NN)~\cite{Feindt:2006pm} 
based on the following input variables:
{\it (1)}~the momentum of the $D_s^+$ in the CM frame.
{\it (2)}~$|dl_{xy}|$ or $|dr|$: $|dl_{xy}|$ is the distance in the $x$-$y$ plane (transverse to the $e^+$ beam) between the $D_s^+$ 
decay vertex and its production vertex. The latter 
is taken to be the $e^+e^-$~IP.
For modes in which there is only one charged track, the $D_s^+$ decay vertex cannot be reconstructed; in this case, we use the variable $|dr|$, which is the impact parameter of the charged track in the $x$-$y$ plane with respect to the IP. 
{\it (3)}\ the cosine of the helicity angle $\theta_h$, which is the angle in the $D_s^+$ rest frame between the momentum of the $K^+$ or $\pi^+$ daughter and the opposite of the boost direction of the
lab frame.
{\it (4)}\ the number of $K^{\pm}$ and $K^0_S$ candidates
reconstructed recoiling against the signal $D_s^+$ candidate. For $e^+e^-\rightarrow c\bar{c}$ events, the charm
quark that does not hadronize to the signal $D_s^+$ 
typically produces a kaon via a $c\rightarrow s$ transition. 
{\it (5)}\ the angle between the $D_s^+$ momentum and 
the thrust axis of the event, both evaluated in the CM frame. 
The thrust axis ($\hat{t}\,$) is defined as the unit vector
that maximizes the quantity 
$\sum_i |\hat{t}\cdot \vec{p}_i|/\sum_i |\vec{p}_i|$, 
where $\vec{p}_i$ are the momenta of particles, and $i$ 
runs over all particles in 
the event. For $e^+e^-\rightarrow c\bar{c}\,$ events,
$D_s^+$ mesons tend to be produced with high momentum, and thus
their direction tends to be close to that of~$\hat{t}$. 
{\it (6)}\ the angle between the $D_s^+$ momentum and 
the vector joining its decay vertex and production vertex
in the $x$-$y$ plane.
This variable is available only for 
$D_s^+\rightarrow (K^+,\pi^+)\eta_{3\pi}$ and
$D_s^+\rightarrow\phi \pi^+$ decays, i.e., modes with 
more than one charged track in the final state.

The NN outputs a single variable ($O^{}_{\rm NN}$), which 
ranges from $-1$ to $+1$. Events with values close to $+1$ ($-1$)
are more signal-like (background-like). For each signal mode,
we require that $O^{}_{\rm NN}$ be greater than some minimum value,
which is determined by optimizing a figure-of-merit (FOM). 
The FOM is taken to be the ratio
$N_{\rm sig}/\sqrt{N_{\rm sig} + N_{\rm bkg}}$, where $N_{\rm sig}$ 
and $N_{\rm bkg}$ are the expected yields of signal and background
events, respectively. 
{The former is evaluated via MC simulation, using
world-average values of branching fractions for signal modes~\cite{PDG2020}}. The latter is evaluated by scaling 
events in data that are reconstructed in a 
mass sideband. This sideband is defined as 
$2.04~{\rm GeV}/c^2 < M_{D_s^+} < 2.10~{\rm GeV}/c^2$ 
for $D_s^+ \rightarrow (K^+, \pi^+)\pi^0$ and
$D_s^+ \rightarrow (K^+, \pi^+)\eta_{\gamma\gamma}$; 
$2.02~{\rm GeV}/c^2 < M_{D_s^+} < 2.05~{\rm GeV}/c^2$ 
for $D_s^+ \rightarrow K^+ \eta_{3\pi}$; and 
$2.02~{\rm GeV}/c^2 < M_{D_s^+} < 2.10~{\rm GeV}/c^2$ 
for $D_s^+ \rightarrow \pi^+ \eta_{3\pi}$. 
For $D_s^+\rightarrow\pi^+\pi^0$ decays, the branching fraction
is unknown; thus, for this mode the FOM is taken to be
$\varepsilon_{\rm sig}/\sqrt{N_{\rm bkg}}$, where 
$\varepsilon_{\rm sig}$ is the reconstruction
efficiency~\cite{Punzi}.
The final selection criteria range from $O^{}_{\rm NN}>0.70$ for $D_s^+\rightarrow\pi^+\eta_{3\pi}$ to $O^{}_{\rm NN}>0.94$ for 
$D_s^+\rightarrow\pi^+\pi^0$.
The corresponding signal efficiencies
range from 35\% for $D_s^+\rightarrow \pi^+\pi^0$ to 63\% for $D_s^+\rightarrow\pi^+\eta_{3\pi}$.

A small fraction of events have multiple $D_s^+$ candidates. 
This fraction ranges from 1\% to 5\%, depending on the decay mode.
For such events, we select one candidate in an event 
by choosing the one with the smallest $\chi^2$ resulting
from the mass-constrained fit of the $\eta$ or $\pi^0$ decay. 
If, after this selection, there are still multiple candidates,
we choose the one with the highest value of~$O^{}_{\rm NN}$.
For the reference mode $D_s^+\rightarrow\phi\pi^+$, which 
has no $\eta$ or $\pi^0$ in the final state, we choose 
the candidate with the highest~$O^{}_{\rm NN}$. 
The efficiency of this best-candidate selection is evaluated
from MC simulation to be about~70\%.

The number of signal events is obtained from an unbinned 
maximum likelihood fit to the $D_s^{+}$ mass distribution. 
For each mode, we perform a simultaneous fit to the $M_{D_s^+}$ 
distributions of both the tagged and untagged samples.  
The nominal fitting range is 1.86--2.07~GeV/$c^2$. However,
for $D_s^+ \rightarrow K^+ (\eta_{\gamma\gamma}, \eta_{3\pi})$ 
and $D_s^+ \rightarrow \pi^+ \eta_{3\pi}$, the range is 
1.86--2.05~GeV/$c^2$.
We fit the $D_s^+$ and $D_s^-$ samples separately but simultaneously.

The following probability density functions (PDFs)
are used for fitting signal and background components. 
For the signal component, the sum of a Crystal Ball (CB) 
function~\cite{CrystalBall} and a Gaussian function,
with both having the same mean, is used.
For $D^+_s\rightarrow \pi^+ \eta_{\gamma\gamma}$ and 
$D^+_s\rightarrow \pi^+ \eta_{3\pi}$, which have high
statistics, the common mean and 
the widths are floated. For other signal modes, 
the means are fixed to those from $D^+_s\rightarrow \pi^+ \eta$,
while the widths are fixed to MC {{simulation}} 
values that are scaled to account for differences in resolution 
between data and the MC.
The scaling factors are determined by comparing signal shape
parameters between data and MC {{simulation}} for $D^+_s\rightarrow \pi^+ \eta$.
The relative fraction of the Gaussian function and two remaining
parameters of the CB function are fixed to MC simulation values.

The dominant background is combinatorial, for which a 
second-order Chebyshev polynomial is used. All background
parameters are floated. The decays 
$D^+ \rightarrow (K^+,\pi^+)\,\pi^0$ and 
$D^+ \rightarrow (K^+,\pi^+)\,\eta$
make peaks in the 
$D_s^+ \rightarrow (K^+,\pi^+)\,\pi^0$ and 
$D_s^+ \rightarrow (K^+,\pi^+)\,\eta$ mass 
distributions; these peaks are described by Gaussian functions.
The $D^+\rightarrow \pi^+\pi^0$ and
$D^+\rightarrow\pi^+\eta$ decays also make peaks in the
$D_s^+ \rightarrow K^+ \pi^0$ and $D_s^+ \rightarrow K^+ \eta$
mass distributions (albeit very small) when the $\pi^{+}$ is
misidentified as a~$K^{+}$. 
The shape of this background and the fractions of
$D^+ \rightarrow \pi^+ \pi^0$ and $D^+ \rightarrow \pi^+ \eta$
decays that are misidentified are taken from MC simulation. 
The yields of 
$D^+ \rightarrow \pi^+ \pi^0$ and $D^+ \rightarrow \pi^+ \eta$
are obtained from the fits to the $\pi^+ \pi^0$ and $\pi^+ \eta$ 
mass distributions.

For the reference mode $D_s^+ \rightarrow \phi \pi^+$, the signal 
PDF is the sum of a bifurcated 
Student's t-distribution~\cite{James_book}
and a Gaussian function. The mean and width of the signal peak 
and the fraction of the Gaussian function are floated.
There is a small background from
$D_s^+ \rightarrow K^+ K^- \pi^+$, in which
the kaons do not originate from $\phi\rightarrow K^+K^-$.
As this background has the same mass distribution as
$D_s^+ \rightarrow \phi \pi^+$, it cannot be distinguished
from {the latter} in the fit. 
We thus correct the $\phi \pi^+$ yield
to account for the $K^+ K^- \pi^+$ contribution.
This contribution is estimated from MC simulation
to be~$(1.73\pm 0.03)$\%~\cite{Zupanc:2013byn}.

\begin{figure}
\includegraphics[width=0.45\textwidth]{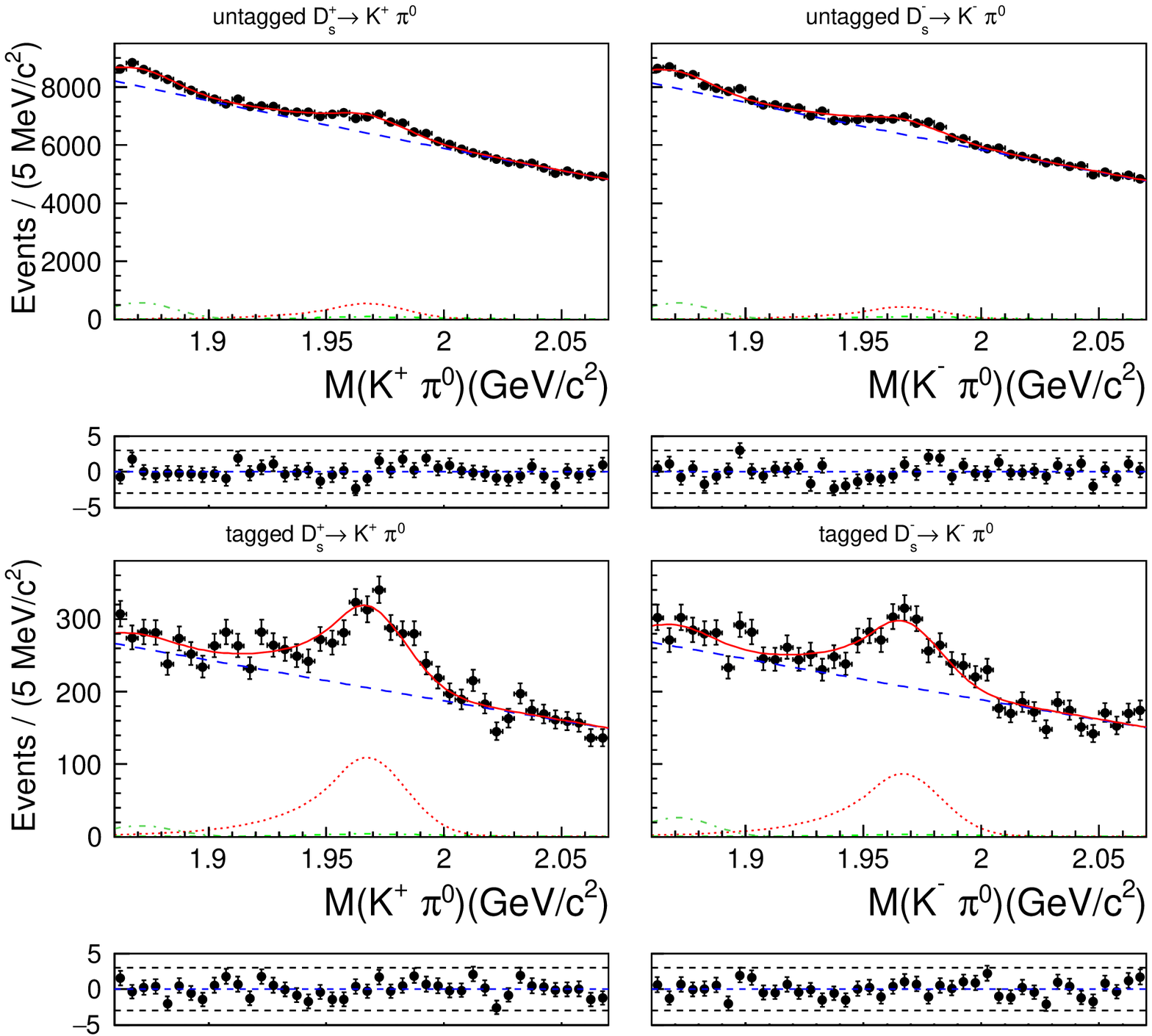}
\includegraphics[width=0.45\textwidth]{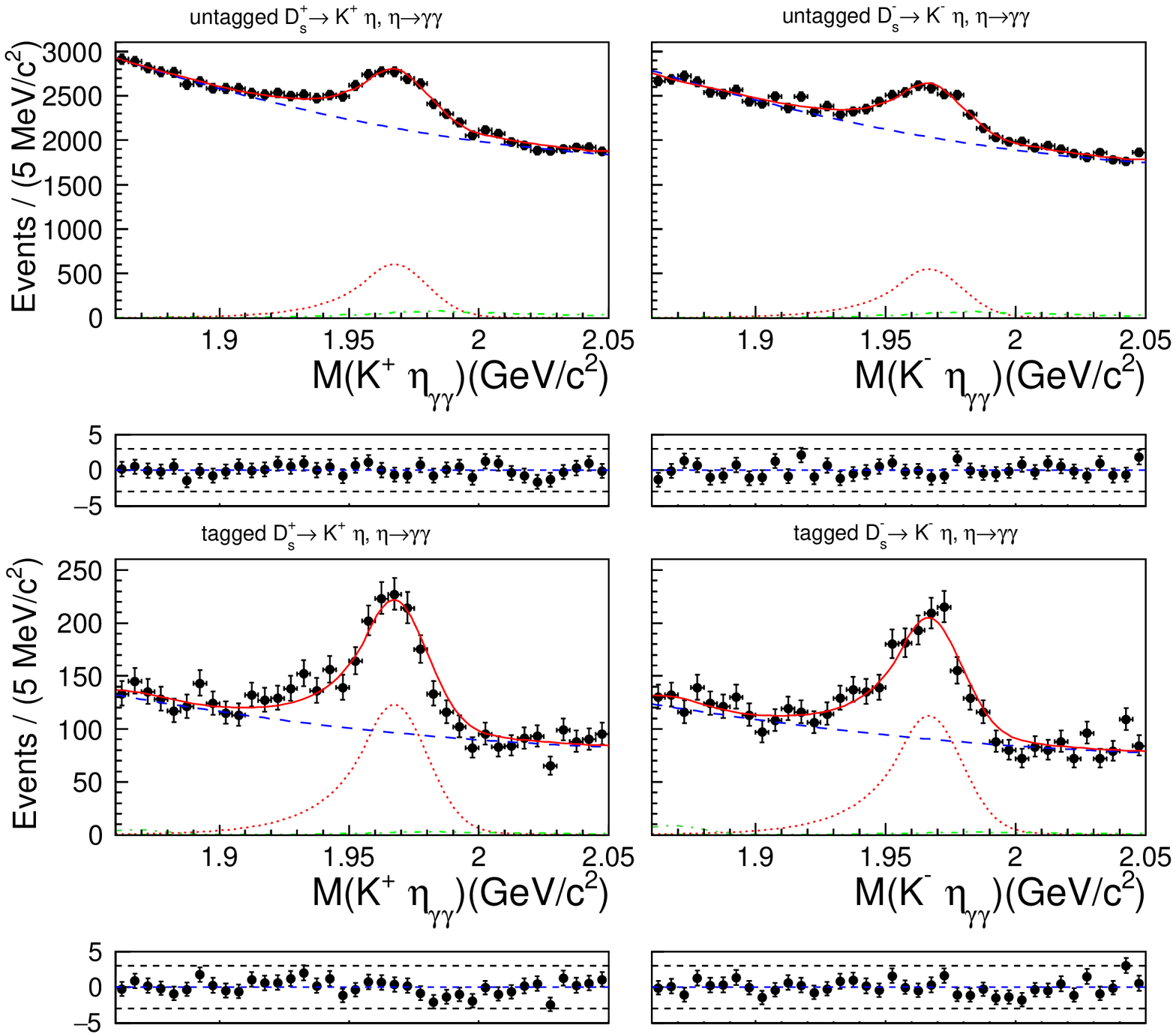}
\includegraphics[width=0.45\textwidth]{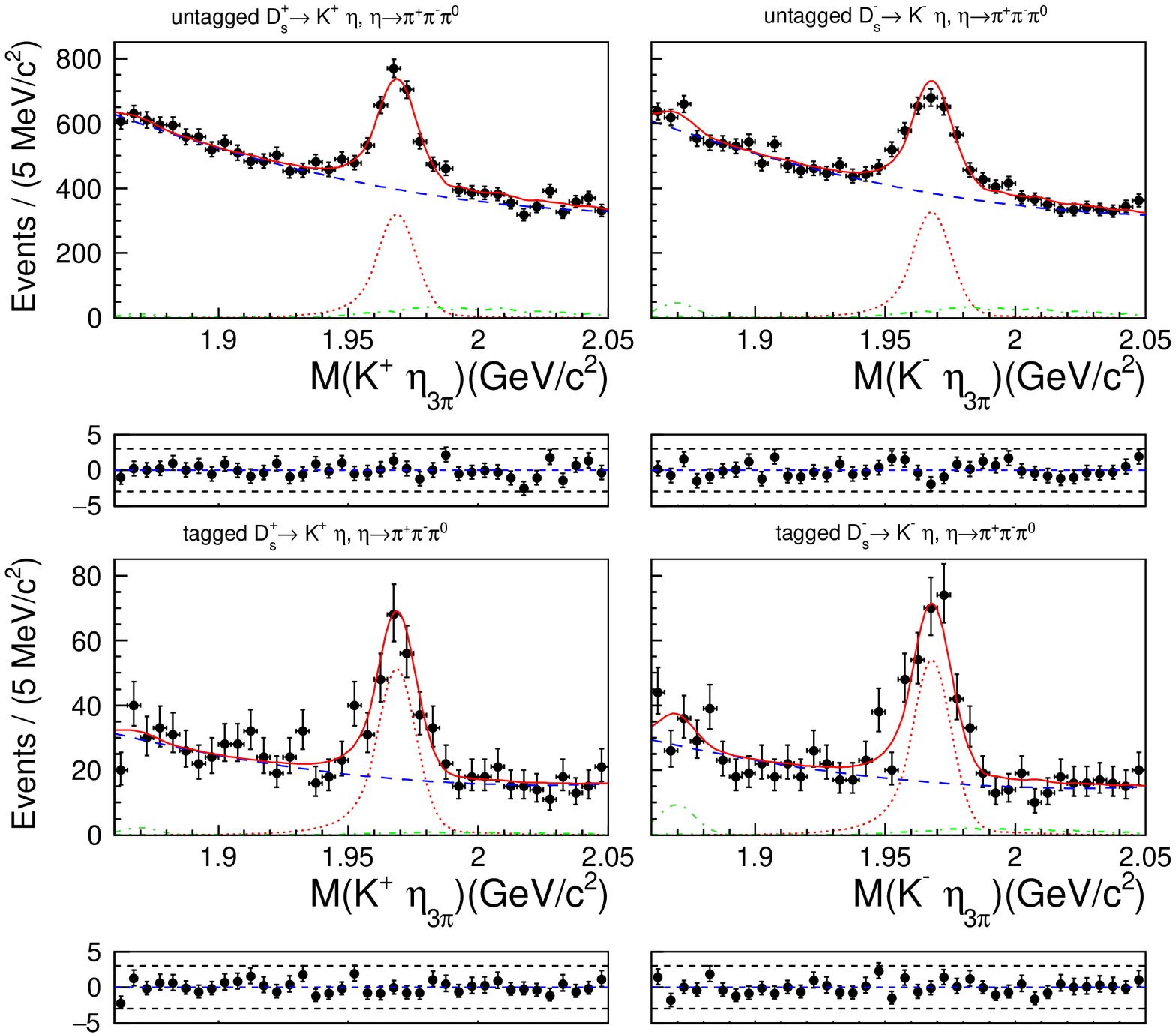}
\caption{(color online) Data and fit projection for $D_s^+ \rightarrow K^+ \pi^0$ (top two rows), 
$D_s^+ \rightarrow K^+ \eta_{\gamma\gamma}$ (middle two rows) and $D_s^+ \rightarrow K^+ \eta_{3\pi}$ (bottom two rows). Left side shows $D_s^+$ candidates, right side shows $D_s^-$ 
candidates. For each pair of rows, top is the untagged sample, bottom is the tagged sample. 
The red solid line is the total fit, the red dotted line is signal, the broken green line is background from $D^+$, and the dashed blue line is combinatorial background. 
The plots beneath the distributions show the residuals.}
\label{fig:fit_kaon}
\end{figure}

\begin{figure}
\includegraphics[width=0.45\textwidth]{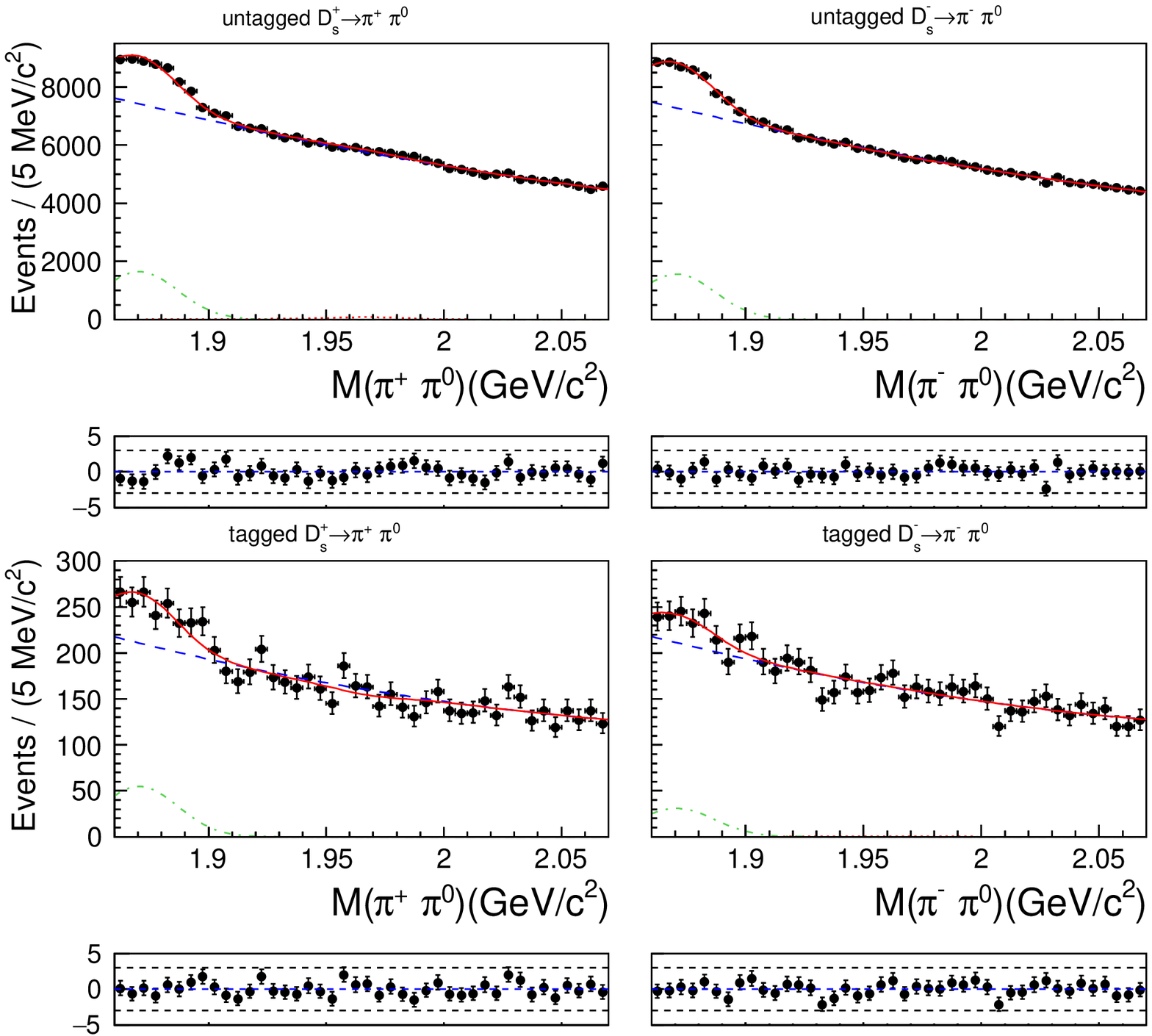}
\includegraphics[width=0.45\textwidth]{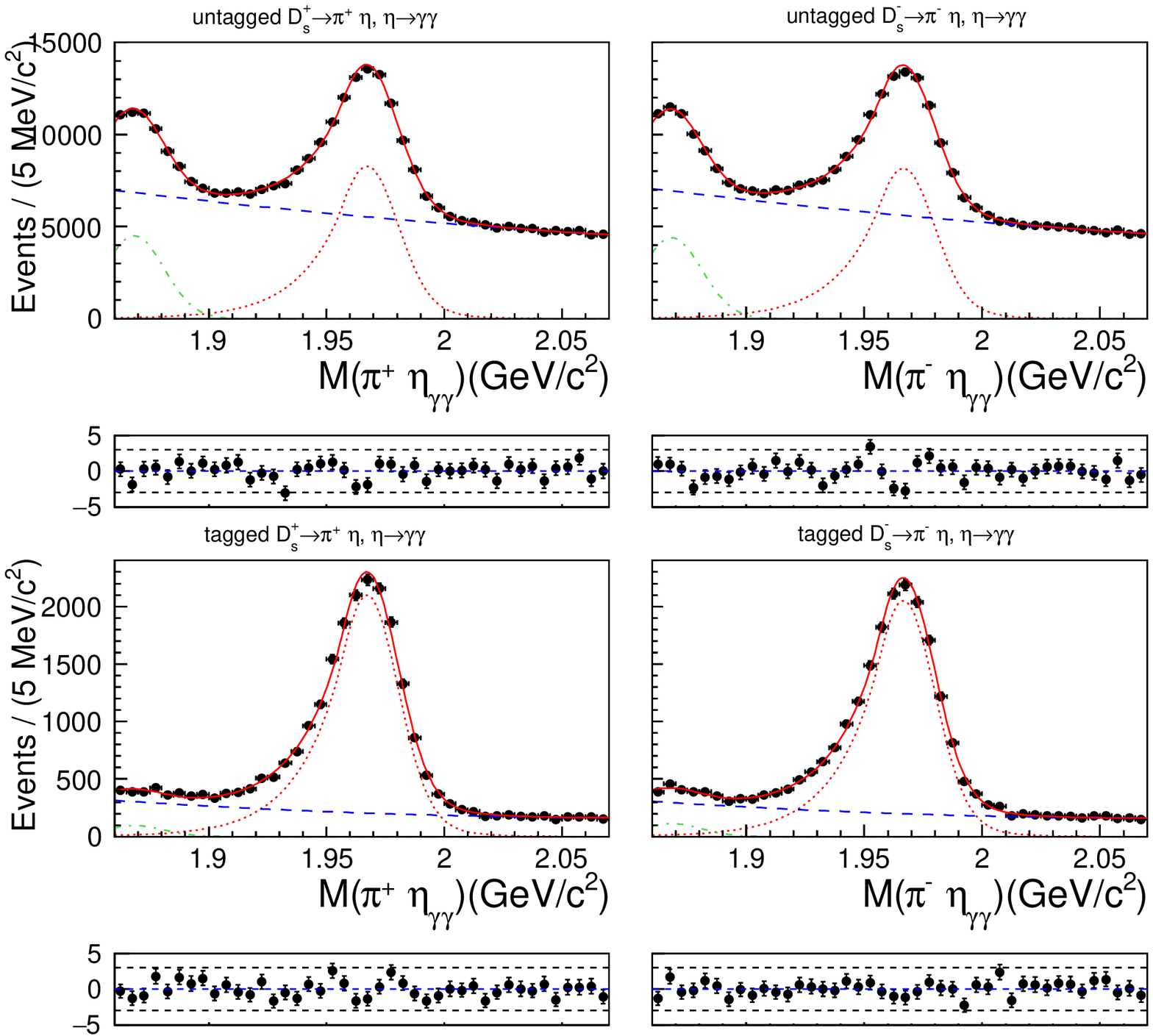}
\includegraphics[width=0.45\textwidth]{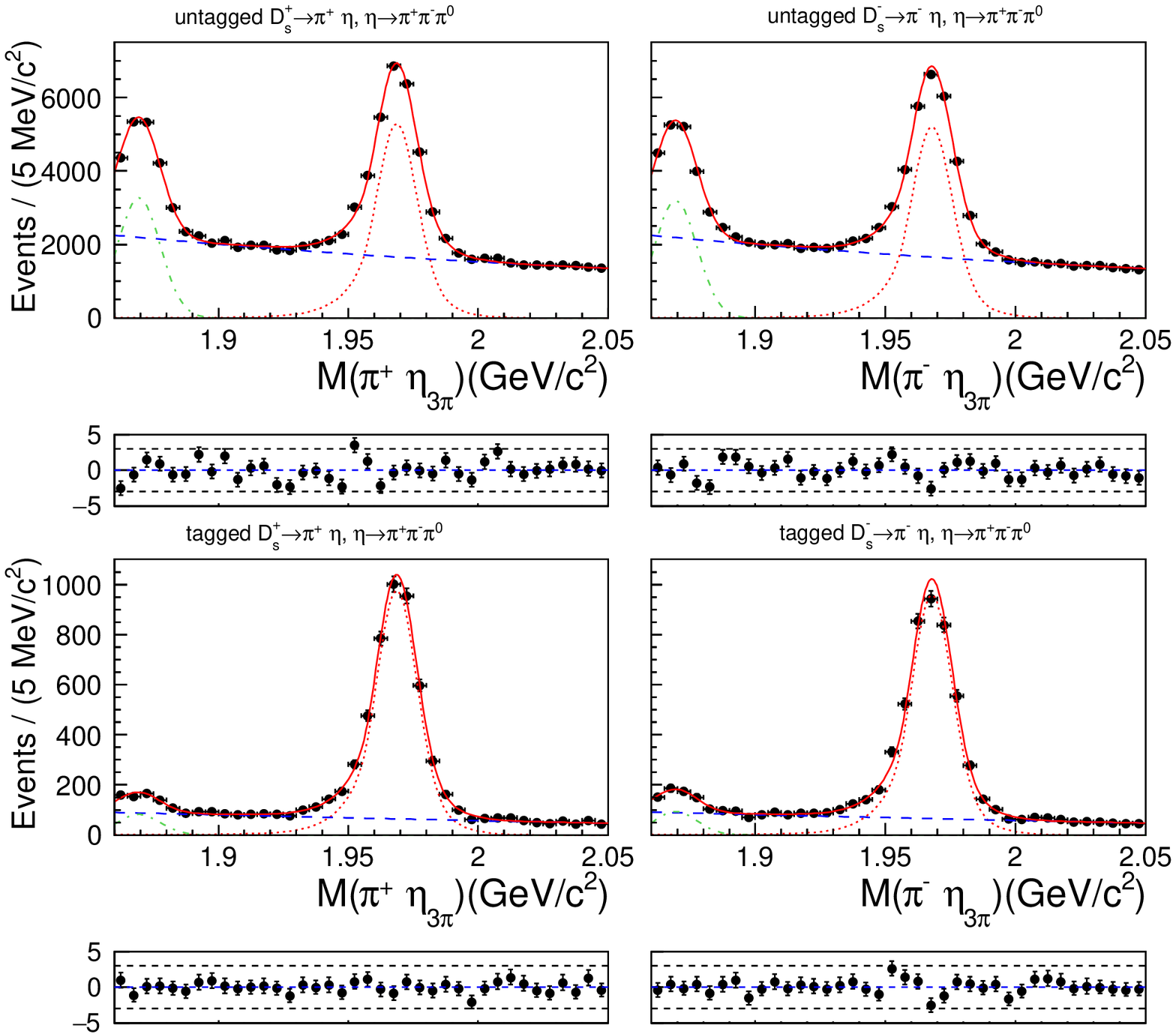}
\caption{(color online) Data and fit projection for $D_s^+ \rightarrow \pi^+ \pi^0$ 
(top two rows), $D_s^+ \rightarrow \pi^+ \eta_{\gamma\gamma}$ (middle two rows) 
and $D_s^+ \rightarrow \pi^+ \eta_{3\pi}$ (bottom two rows). 
Left side shows $D_s^+$ candidates, right side shows $D_s^-$ 
candidates. For each pair of rows, top is the untagged sample, bottom is the tagged sample. 
The red solid line is the total fit, the red dotted line is signal, the broken green line is background from $D^+$, and the dashed blue line is combinatorial background. 
The plots beneath the distributions show the residuals.}
\label{fig:fit_pion}
\end{figure}

\begin{figure}
\includegraphics[width=0.45\textwidth]{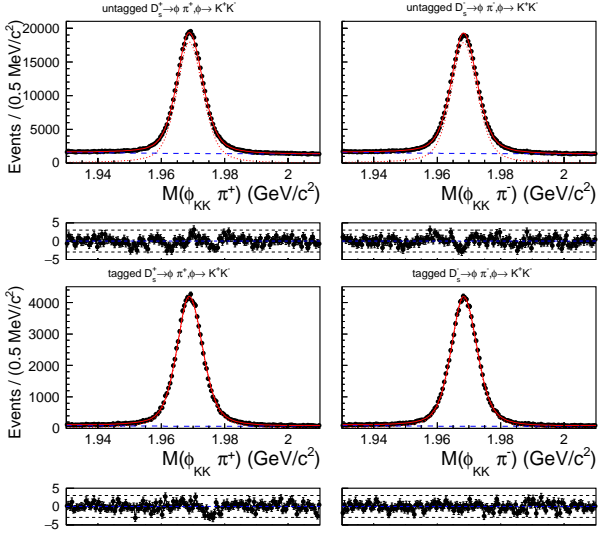}
\caption{(color online) Data and fit projection for the reference mode
$D_s^+ \rightarrow \phi \pi^+$. 
Left side shows $D_s^+$ candidates, right side shows $D_s^-$ 
candidates. Top row is the untagged sample, bottom row is the tagged sample. 
The red solid line is the total fit, the red dotted line is signal, and the 
dashed blue line is background. 
The plots beneath the distributions show the residuals.}
\label{fig:fit_phipi}
\end{figure}

The $M_{D_s^+}$ distributions along with projections of the fit result
are shown in Figs.~\ref{fig:fit_kaon}, \ref{fig:fit_pion}, and~\ref{fig:fit_phipi}. 
The branching fraction ${\cal B}_{\rm sig}$ for
the signal modes is calculated as
\begin{equation}
{\cal B}_{\rm sig} = 
\left(\frac{N_{\rm sig}}{N_{\phi\pi^+}}\right)  \left(\frac{\varepsilon_{\phi\pi^+}}{\varepsilon_{\rm sig}}\right)
\cdot {\cal B}_{\phi\pi^+} \,,
\end{equation}
where $N_{\rm sig}$ and $N_{\phi\pi^+}$ are the yields of the 
signal and reference mode $D_s^+\rightarrow\phi\pi^+$, respectively.
Each yield is the sum of the yields for the 
tagged and untagged samples. 
The terms $\varepsilon_{\rm sig}$ 
and $\varepsilon_{\phi\pi^+}$ are
the corresponding reconstruction efficiencies,
as evaluated from MC simulation.
The branching fraction ${\cal B}_{\phi\pi^+}$ for 
$D_s^+\rightarrow\phi (\rightarrow K^+ K^-)\pi^+$ is taken to be the
world-average value $(2.24\pm 0.08)$\%~\cite{PDG2020}. 

All signal yields and resulting branching fractions are listed 
in Table~\ref{table:BRmea_data}.
A weighted average of the results from the two $\eta$ decay 
channels ($\eta_{\gamma\gamma}$ and $\eta_{3\pi}$) is also given.
The results listed include systematic uncertainties, which are
discussed later.
As no significant signal for $D_s^+\rightarrow\pi^+\pi^0$
is observed, we set an upper limit on its branching fraction 
using a Bayesian approach. We calculate the likelihood function
${\cal L}$ as a function of branching fraction; 
the value $\xi$ that satisfies $\int_0^{\xi} {\cal L}(x)\,dx=0.90$
is taken to be the 90\% confidence level ({C.L.}) upper limit. 
{We include systematic uncertainty into this limit by convolving
${\cal L}(x)$, before integrating, with a Gaussian function 
whose width is equal to the total systematic uncertainty.
}
The result is 
${\cal B}(D_s^+ \rightarrow \pi^+ \pi^0) < 1.2 \times 10^{-4}$ 
at 90\% {C.L}.

\begin{table*}
\renewcommand{\arraystretch}{1.2}
\begin{center}
\caption{Reconstruction efficiencies, fitted signal yields, and resulting relative and absolute branching fractions. 
The yields listed are the sums of those from the tagged and 
untagged samples. The first and second uncertainties listed are statistical and systematic, respectively.
The third uncertainty is due to
the external branching fraction ~${\cal B}^{}_{\phi\pi^+}$. Results from the two $\eta$ 
decay modes are combined via a weighted average and also listed.
{All results are corrected for the
$\pi^0\rightarrow \gamma\gamma$, $\eta\rightarrow \gamma\gamma$, 
or $\eta\rightarrow \pi^+\pi^-\pi^0$ branching fractions.}}
\label{table:BRmea_data}
\begin{tabular}{lcccccc}
\hline
\hline
Decay mode & $\varepsilon$ (\%) & Fitted yield &   ${\cal B}/{\cal B}_{\phi\pi^+}$ (\%) &  ${\cal B}$ ($10^{-3}$)  \\
\hline
$D_s^+ \rightarrow K^+ \pi^0$  & 
$8.10 \pm 0.04$    & 11978 $\pm$ 846     &  3.28 $\pm$ 0.23 $\pm$ 0.13  & 0.735 $\pm$ 0.052 $\pm$ 0.030 $\pm$ 0.026   \\
$D_s^+ \rightarrow K^+ \eta_{\gamma\gamma}$ & 
$7.42 \pm 0.05$  &  10716 $\pm$ 429   &  8.04 $\pm$ 0.32 $\pm$ 0.35     & 1.80 $\pm$ 0.07 $\pm$ 0.08 $\pm$ 0.06  \\
$D_s^+ \rightarrow K^+ \eta_{3\pi}$  & 
$4.04 \pm 0.02$  &  3175  $\pm$  121   &  7.62 $\pm$ 0.29 $\pm$ 0.33    & 1.71 $\pm$ 0.07  $\pm$ 0.08 $\pm$ 0.06     \\
$D_s^+ \rightarrow K^+ \eta$ & $-$ &  $-$  & 7.81 $\pm$ 0.22 $\pm$ 0.24 & 1.75 $\pm$ 0.05 $\pm$  0.05 $\pm$ 0.06  \\
\hline
$D_s^+ \rightarrow \pi^+ \pi^0$ & 
$6.63 \pm 0.04$  &  491 $\pm$ 734        &  0.16 $\pm$ 0.25 $\pm$ 0.09    & 0.037 $\pm$ 0.055  $\pm$ 0.021 $\pm$ 0.001   \\
$D_s^+ \rightarrow \pi^+ \eta_{\gamma\gamma}$ 
&  $10.84 \pm 0.02$ & 166696 $\pm$ 1173  & 85.54 $\pm$ 0.64 $\pm$ 3.32  & 19.16 $\pm$ 0.14 $\pm$ 0.74 $\pm$ 0.68 \\
$D_s^+ \rightarrow \pi^+ \eta_{3\pi}$ 
& $6.50 \pm 0.03$     &  56132 $\pm$ 407     & 83.55 $\pm$ 0.64 $\pm$ 4.37  & 18.72 $\pm$ 0.14 $\pm$ 0.98 $\pm$ 0.67 \\
$D_s^+ \rightarrow \pi^+ \eta$ & $-$ & $-$ & 84.80 $\pm$ 0.47 $\pm$ 2.64 & 19.00 $\pm$  0.10 $\pm$  0.59 $\pm$  0.68  \\
\hline
 $D_s^+ \rightarrow \phi \pi^+$  
 & $22.05 \pm 0.13$       &  1005688 $\pm$ 2527  &   1  & $-$  \\
\hline
\hline
\end{tabular}
\end{center}
\end{table*}

As the $D_s^+$ and $D_s^-$ samples are fitted separately, 
we obtain the raw asymmetry $A_{\rm{raw}}$, defined as
 \begin{equation}
A_{\rm{raw}} = 
\frac{N_{D_s^+} - N_{D_s^-} }{N_{D_s^+} + N_{D_s^-}}\,.
\end{equation} 
In this expression, $N_{D_s^+}$ ($N_{D_s^-}$) is the 
signal yield for the $D_s^+$ ($D_s^-$) sample.
This raw asymmetry receives three contributions:
\begin{equation}
A_{\rm{raw}} = A_{CP} + A_{\rm FB} + A_{\epsilon}\,,
\end{equation}
where $A_{CP}$ is the intrinsic $CP$ asymmetry of interest; 
$A_{\rm FB}$ is the ``forward-backward'' asymmetry that arises 
from interference between amplitudes mediated by a virtual 
photon and by a $Z^0$ boson;
and $A_{\epsilon}$ is an asymmetry that arises from a 
possible difference in reconstruction efficiencies between 
positively charged and negatively charged tracks.
The asymmetry $A_{\rm FB}$ is an odd function of the cosine 
of the $D_s^{+}$ polar angle in the CM frame 
($\cos\theta^{\rm CM}_{D_s}$). 
The asymmetry $A_{\epsilon}$ arises from small differences
in tracking and particle identification efficiencies and depends 
on the momentum and polar angle of the charged track.
For $D_s^{*+}\rightarrow D_s^+\gamma$ decays, 
we find that the momentum distribution of the $\pi^+$ 
or $K^+$ in the $D_s^+$ decay is essentially the 
same as that in prompt $D_s^+$ decays. Thus, 
for a $D_s^+$ decay mode, we take $A_{\epsilon}$ 
to be the same for both the tagged and untagged samples.

For the mode $D_s^+ \rightarrow \pi^+ \eta$,
we correct for $A_{\rm FB}$ and $A_{\epsilon}$ using the reference 
mode $D_s^+ \rightarrow \phi \pi^+$. As the momentum spectrum and
polar angle distributions of the $\pi^+$ daughters in both decay 
modes are essentially identical, the asymmetry $A_{\epsilon}$ is 
expected to be the same. As the asymmetry $A_{\rm FB}$ is 
independent of decay mode, subtracting the
$D_s^+ \rightarrow \pi^+ \eta$ and $D_s^+ \rightarrow \phi \pi^+$
raw asymmetries yields the difference in $CP$ asymmetries:
\begin{eqnarray}
\Delta A_{\rm raw} & \equiv & 
A^{\pi \eta}_{\rm raw}  - A^{\phi \pi}_{\rm raw}\ =\ 
A^{\pi \eta}_{CP} - A^{\phi \pi} _{CP}\,.
\end{eqnarray}
Thus, 
$A^{\pi \eta}_{CP} =  \Delta A_{\rm raw} + A^{\phi \pi} _{CP}$.
Inserting the well-measured value 
$A^{\phi \pi} _{CP} = -0.0038\pm 0.0026\pm 0.0008$~\cite{PDG2020}
subsequently yields~$A^{\pi \eta}_{CP}$.

\begin{figure}
\includegraphics[width=0.23\textwidth]{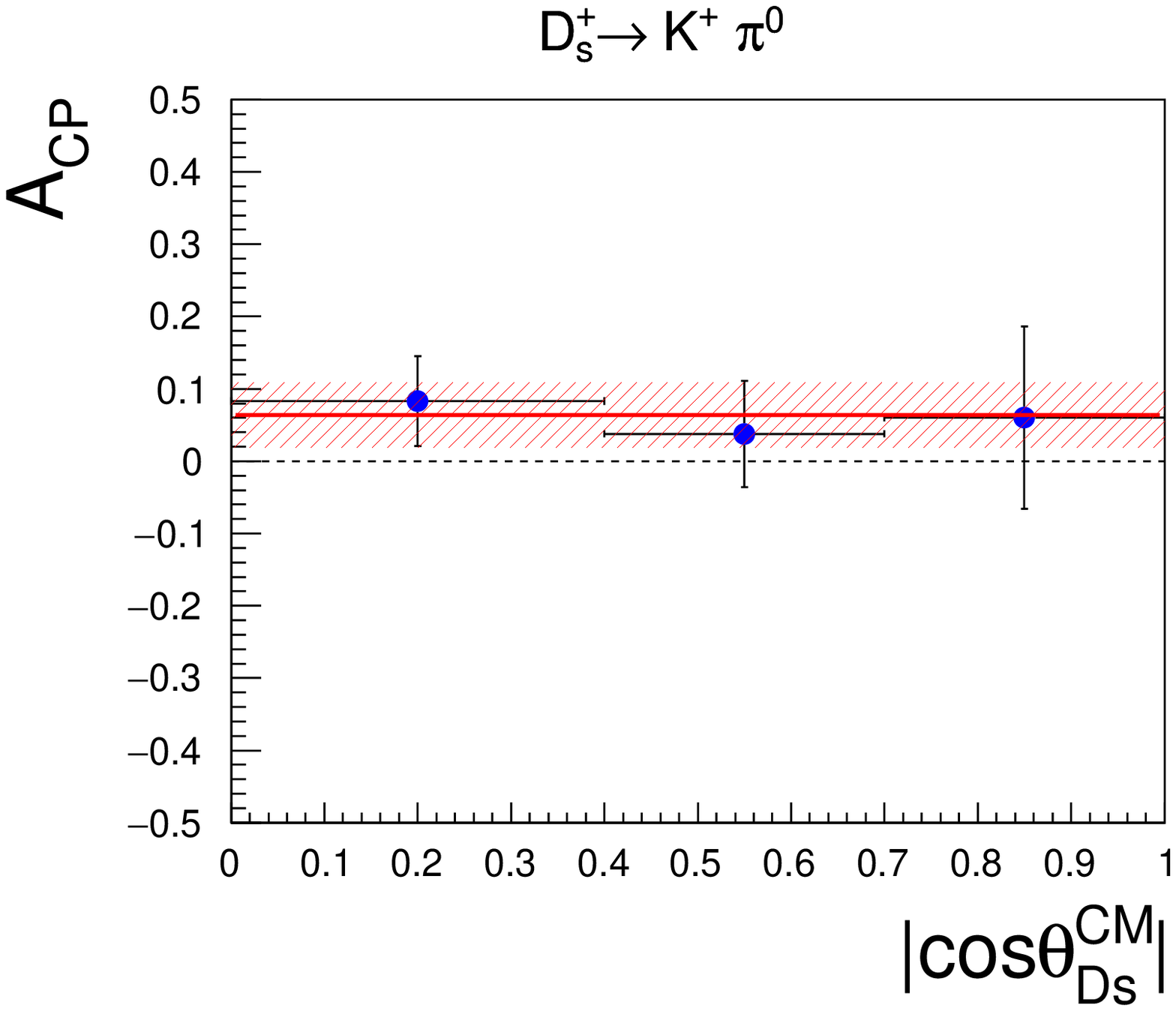}
\includegraphics[width=0.23\textwidth]{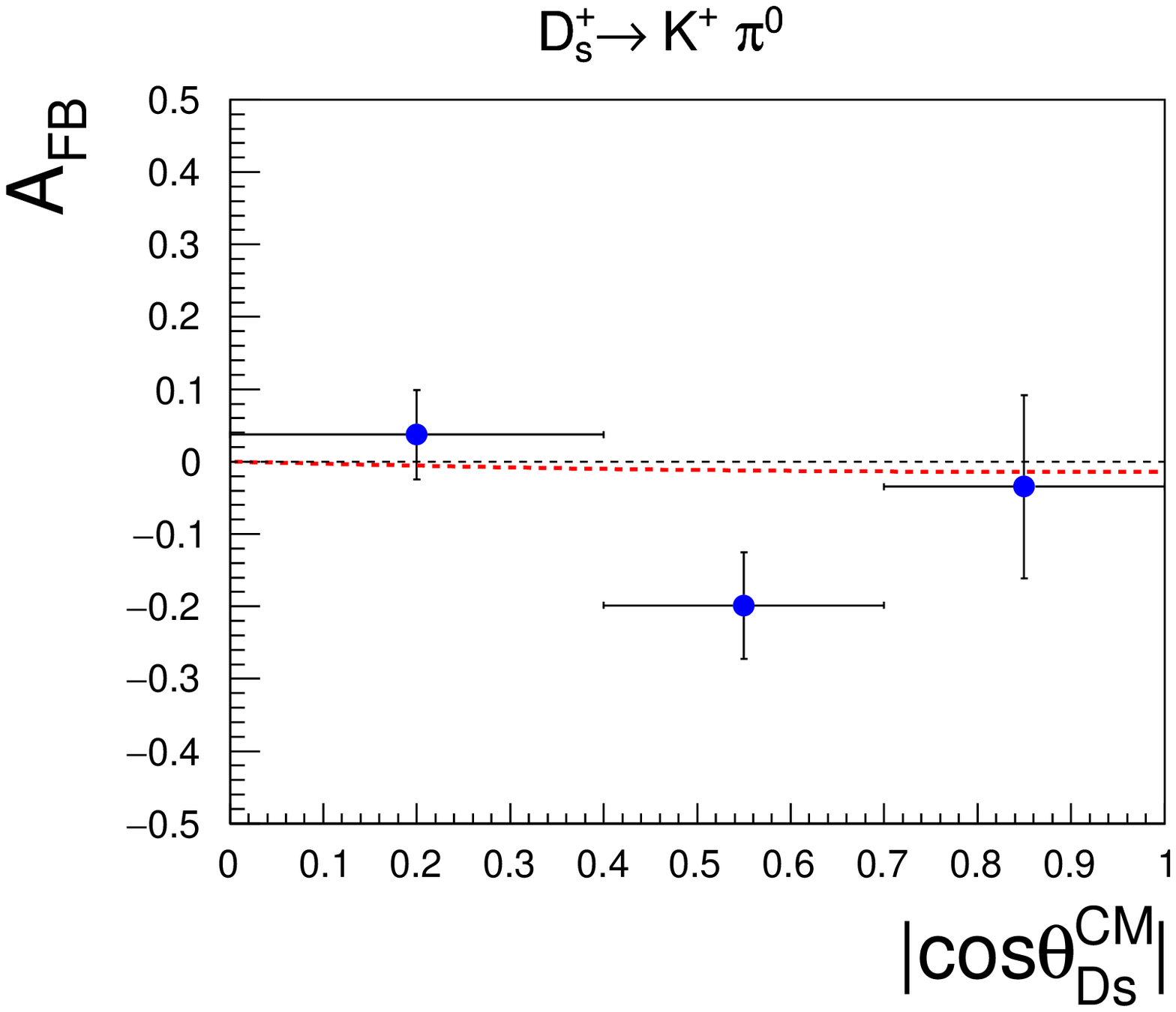}
\includegraphics[width=0.23\textwidth]{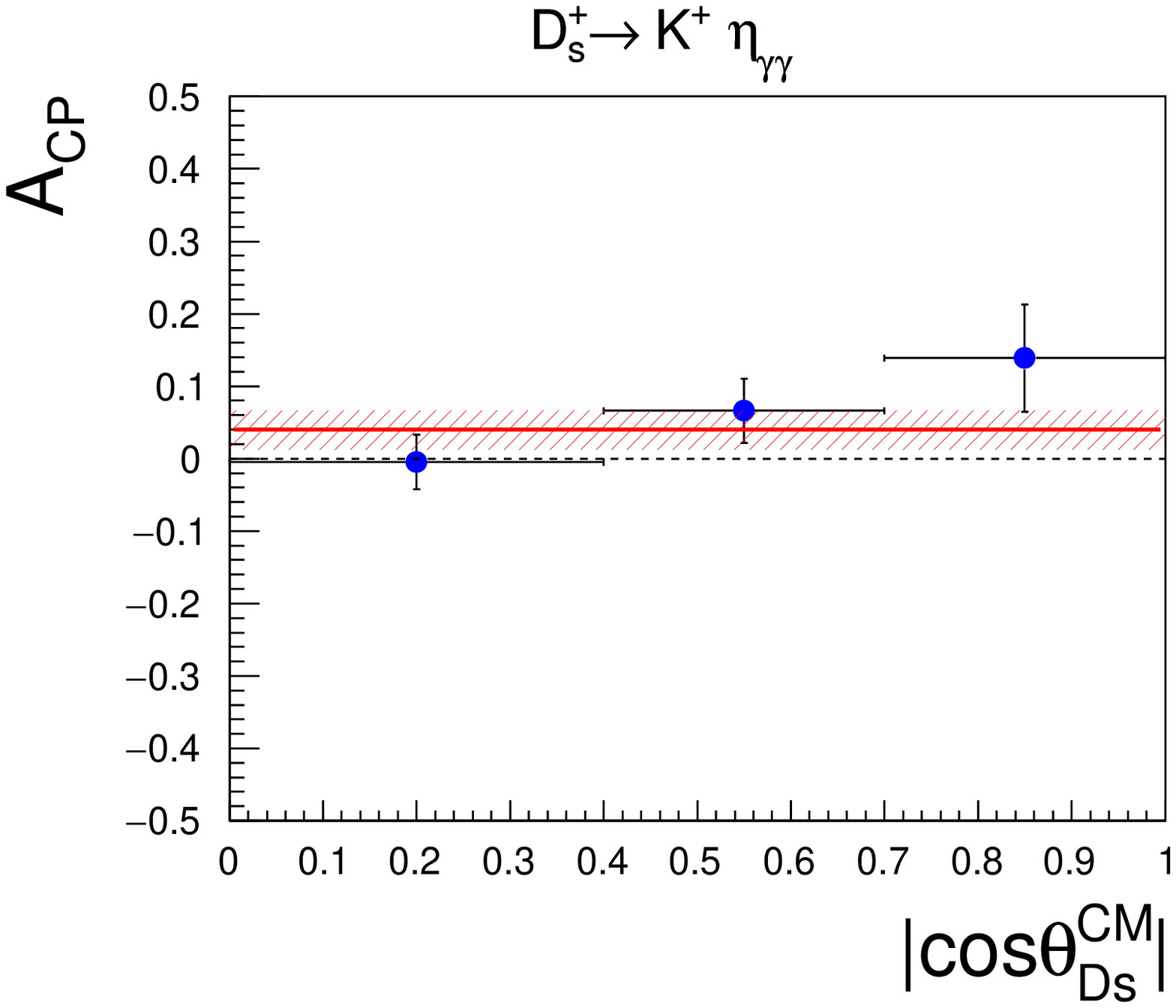}
\includegraphics[width=0.23\textwidth]{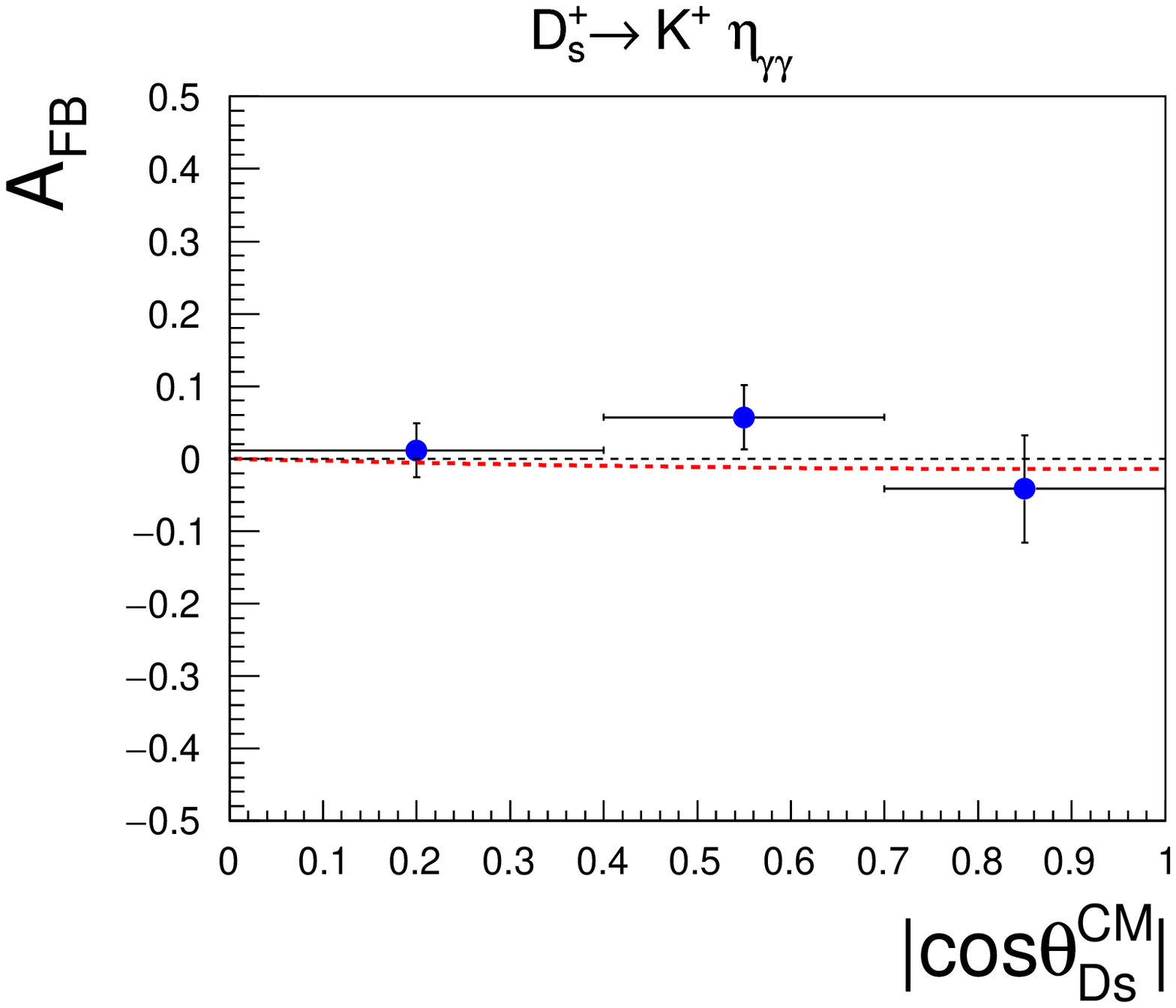}
\includegraphics[width=0.23\textwidth]{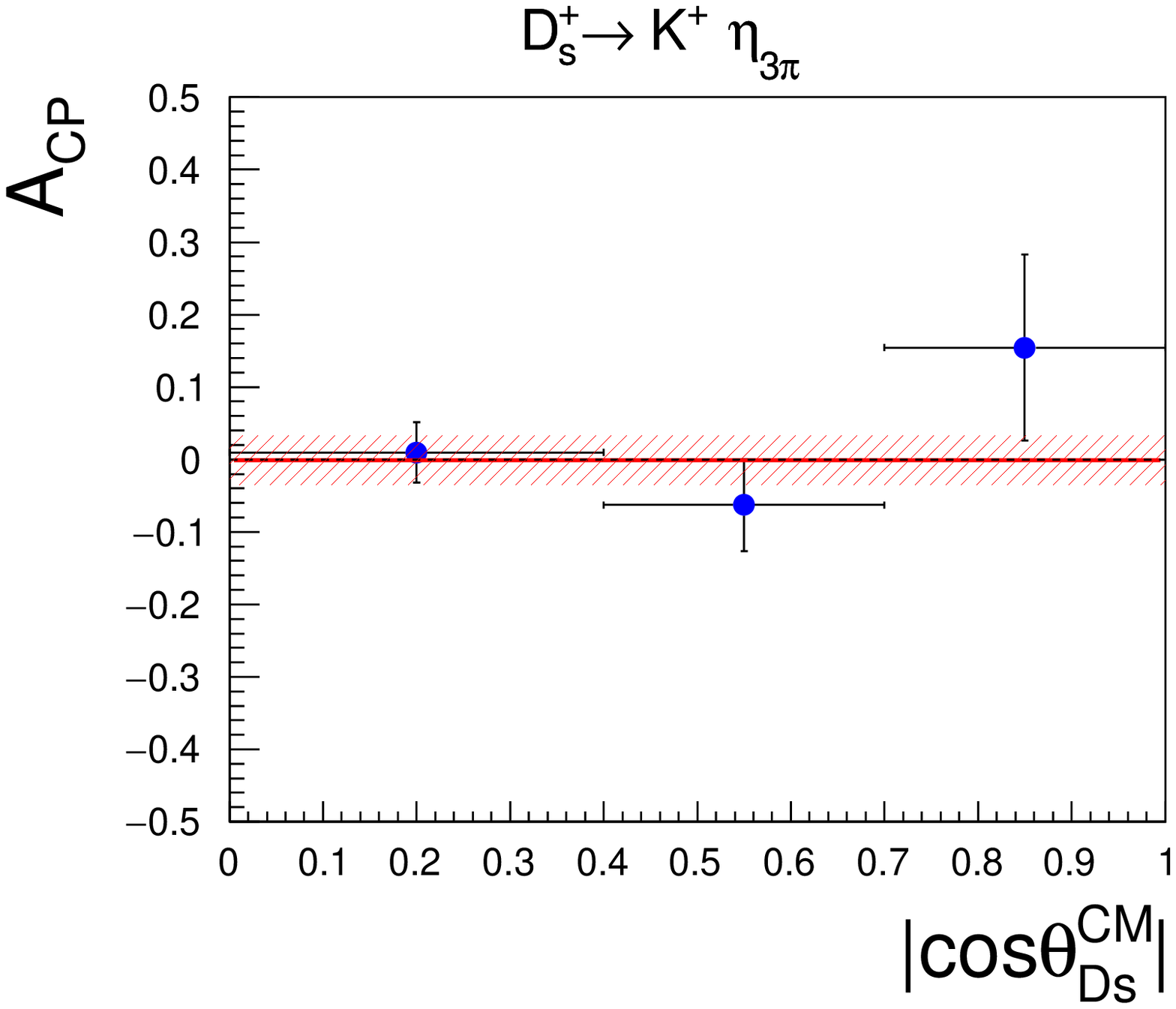}
\includegraphics[width=0.23\textwidth]{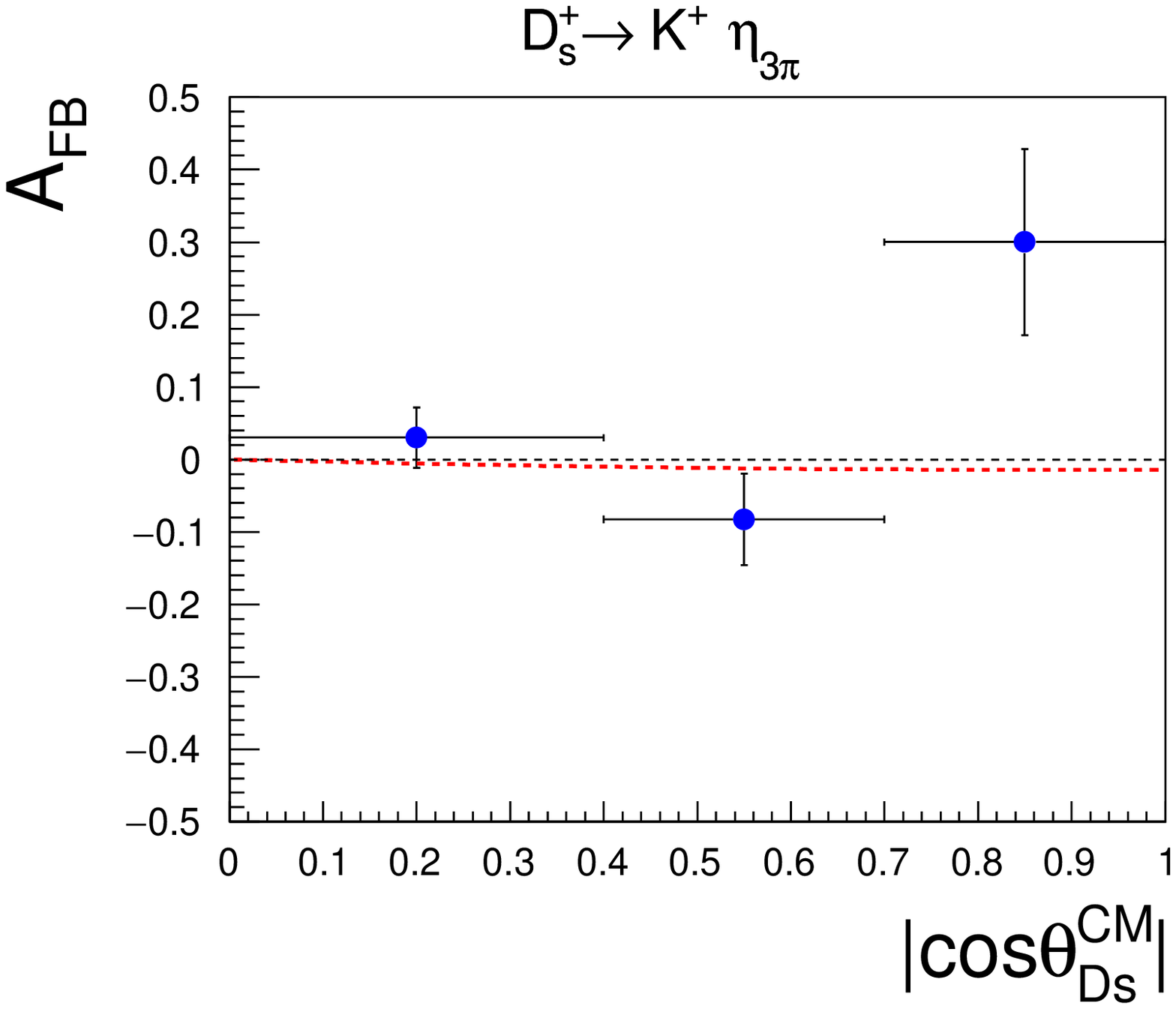}
\caption{(color online) $CP$ asymmetries (left) 
and $A_{\rm FB}$ (right) in bins of 
$|\cos\theta^{\rm CM}_{D_s}|$, 
for $D_s^+ \rightarrow K^+ \pi^0$ (top), 
$D_s^+ \rightarrow K^+ \eta_{\gamma\gamma}$ (middle), and 
$D_s^+ \rightarrow K^+ \eta_{3\pi}$ (bottom). 
In the left-side  plots, the horizontal line shows 
the result of a fit to a constant, and the 
red shaded region shows the $\pm 1\sigma$ errors.
In the right-side plots, the dashed line show the leading-order 
prediction~\cite{AFB}.}
\label{fig:ACP}
\end{figure}

For signal modes $D_s^+ \rightarrow K^+ \pi^0$ and 
$D_s^+ \rightarrow K^+ \eta$, the mode 
$D_s^+ \rightarrow \phi \pi^+$ cannot be used to correct 
for $A_{\epsilon}$ as the daughters are different types.
In this case, we calculate $A_{\epsilon}$ using previous 
Belle measurements of $K^\pm$ efficiencies made as a function 
of track momentum and polar angle~\cite{Staric:2011en}.
We convolve this two-dimensional efficiency map 
with the corresponding momentum and angular distributions,
as determined from MC, of the $K^\pm$ tracks in our signal 
modes to obtain~$A_{\epsilon}$. The resulting values of 
$A_{\epsilon}$ range from $-0.001$ to $-0.008$.
Correcting for this asymmetry results in $A_{\rm corr}$, 
which is the sum of $A_{CP}$ and $A_{\rm FB}$.
As $A_{\rm FB}$ is an odd function of the polar angle 
$\cos\theta^{\rm CM}_{D_s}$, we extract $A_{CP}$ and 
$A_{\rm FB}$ by calculating
\begin{eqnarray}
A_{CP} & = & 
\frac{A_{\rm{corr}}(\cos\theta^{{\rm {CM}}}_{D_s}) +
A_{\rm{corr}}(-\cos\theta^{{\rm{CM}}}_{D_s})}{2}  \,,
\nonumber \\
A_{\rm FB} & = & \frac{A_{\rm{corr}}(\cos\theta^{CM}_{D_s}) - A_{\rm{corr}}(-\cos\theta^{CM}_{D_s})}{2}\,.
\end{eqnarray}
We perform this calculation in six bins of $\cos\theta^{{\rm CM}}_{D_s}$:
$[-1.0, -0.7]$, $[-0.7,-0.4]$, $[-0.4, 0]$, 
$[0, 0.4]$, $[0.4, 0.7]$, and $[0.7, 1.0]$.
The results for  $A_{CP}$ and $A_{\rm FB}$ 
are plotted in Fig.~\ref{fig:ACP}.
We subsequently fit these points to a constant to obtain
final values of $A_{CP}$; the results are listed in 
Table~\ref{table:CPV_data}. 
For $D_s^+ \rightarrow \pi^+ \pi^0$, no signal is 
observed, and thus there is no result for $A_{CP}$.

\begin{table}
\renewcommand{\arraystretch}{1.2}
\begin{center}
\caption{Measured $CP$ asymmetries. 
The first and second uncertainties listed are statistical and systematic, respectively. 
Results from the two $\eta$ decay modes are combined via a weighted average and also listed.}
\label{table:CPV_data}
\begin{tabular}{lccc}
\hline
\hline
Decay mode &   $A_{\rm {raw}}$  & $A_{CP} $  \\
\hline
$D_s^+ \rightarrow K^+ \pi^0$               &   0.115 $\pm$ 0.045   &  0.064  $\pm$ 0.044 $\pm$ 0.011        \\
$D_s^+ \rightarrow K^+ \eta_{\gamma\gamma}$ &   0.046 $\pm$ 0.027   &  0.040   $\pm$ 0.027 $\pm$ 0.005    \\
$D_s^+ \rightarrow K^+ \eta_{3\pi}$         &   $-$0.0112 $\pm$ 0.033 &  $-$0.008  $\pm$ 0.034 $\pm$ 0.008      \\
$D_s^+ \rightarrow K^+ \eta$                &   $-$ & 0.021 $\pm$ 0.021 $\pm$ 0.004    \\
\hline
$D_s^+ \rightarrow \pi^+ \eta_{\gamma\gamma}$   & 0.007 $\pm$ 0.004   &   0.002 $\pm$ 0.004 $\pm$ 0.003  \\
$D_s^+ \rightarrow \pi^+ \eta_{3\pi}$       &   0.008 $\pm$ 0.006    &  0.002 $\pm$ 0.006 $\pm$ 0.003  \\
$D_s^+ \rightarrow \pi^+ \eta$              &  $-$ & 0.002 $\pm$ 0.003 $\pm$ 0.003   \\
\hline
 $D_s^+ \rightarrow \phi\pi^+$              & 0.002 $\pm$ 0.001 & $-$    \\
\hline
\hline
\end{tabular}
\end{center}
\end{table}


The systematic uncertainties for the branching fraction 
are summarized in Table~\ref{table:sys_BR}.
The uncertainty due to charged track reconstruction is 
evaluated from a study of partially reconstructed 
$D^{*+} \rightarrow \pi^+ D^0\,(\rightarrow K_S^0\,\pi^+ \pi^-)$ 
decays and found to be 0.35\% per track. 
The uncertainty due to particle 
identification is evaluated from a study of 
$D^{*+} \rightarrow \pi^+ D^0\,(\rightarrow K^-\pi^+)$ decays.
We note that the uncertainties due to tracking and particle 
identification partially cancel between the signal and reference modes. 
The uncertainty due to  $\pi^0/\eta \rightarrow \gamma\gamma$ 
reconstruction is evaluated from a study of 
$\tau^{-} \rightarrow \pi^- \pi^0\,\nu_\tau$ 
decays and found to be~2.4\%.

To study the systematic uncertainty due to the $O^{}_{\rm NN}$ 
requirement, we remove this requirement for the
high-statistics $D_s^+ \rightarrow \pi^+ \eta$ mode
and also for the reference mode $D_s^+ \rightarrow \phi\pi^+$. 
We subsequently use the ${_{s}\mathcal{P}lot}$~\cite{Pivk:2004ty} 
technique to extract
the $O^{}_{\rm NN}$ distribution for each decay.
From these distributions, we calculate the 
efficiencies of the $O^{}_{\rm NN}$ requirements used 
for the six signal decay modes.
We repeat this calculation for both 
data and MC samples and take the difference between the resulting 
efficiencies as the systematic uncertainty due to the 
$O^{}_{\rm NN}$ requirement. This uncertainty ranges from 
0.9\% to 1.2\% for the signal modes, and is 0.6\% 
for the reference mode.

There is systematic uncertainty in the reconstruction efficiencies
$\varepsilon^{}_{\rm sig}$ and $\varepsilon^{}_{\phi\pi^+}$ 
arising from a possible difference between MC and data 
in the fraction of $D_s^+$ decays originating from 
$D_s^{+*}\rightarrow D_s^+\gamma$. This difference 
is common to both signal and normalization modes
and nominally cancels out in the ratio
$\varepsilon^{}_{\phi\pi^+}/\varepsilon^{}_{\rm sig}$.
However, there could be a small difference remaining 
if there were a difference in reconstruction efficiencies 
between tagged and untagged $D_s^+$ decays, and this difference
differed between signal and normalization modes. 
Thus, the systematic uncertainty in the ratio
$\varepsilon^{}_{\phi\pi^+}/\varepsilon^{}_{\rm sig}$
due to such differences is found to be small, only 0.7\%. 
The statistical errors on 
$\varepsilon^{}_{\rm sig}$ and $\varepsilon^{}_{\phi\pi^+}$
due to the limited sizes of the 
MC samples used to evaluate them are taken as a
systematic uncertainty.

The systematic uncertainties due to the fitting procedure 
are as follows.
{\it (a)\/}
The uncertainty due to fixed parameters in 
the fits is estimated by varying these parameters according 
to their uncertainties. For each signal mode, we vary all 
such parameters simultaneously, repeating the fit 1000 times. 
We plot the fit results and take the r.m.s. 
of these distributions as the systematic uncertainty.
{\it (b)\/}
The uncertainty due to the amount of peaking 
background from $D^+\rightarrow \pi^+ (\pi^0/\eta)$ decays
is evaluated by varying this background by~$\pm 1\sigma$; 
the resulting changes in the 
signal yields are assigned as systematic uncertainties. 
{\it (c)\/}
The uncertainty due to the choice of fitting range is evaluated
by varying this range; the change in the branching fraction
is assigned as a systematic uncertainty.
{\it (d)\/} To evaluate potential fit bias, we perform 1000 fits to ``toy" MC samples. Small possible differences observed 
between the fitted signal yields and the input values are 
assigned as systematic uncertainties.

The uncertainty on the branching fraction for the 
reference mode $D_s^+ \rightarrow \phi \pi^+$,
which is taken from 
Ref.~\cite{PDG2020} and is external to the analysis,
is taken as a systematic uncertainty. All uncertainties are 
added in quadrature to
give, for each signal mode, an overall systematic 
uncertainty. These overall uncertainties are also listed 
in Table~\ref{table:sys_BR}.

\begin{table*}
\renewcommand{\arraystretch}{1.2}
\begin{center}
\caption{Systematic uncertainties for the ratio of branching fractions, in percent. 
The overall uncertainty is the sum in quadrature of the listed uncertainties
and corresponds to the systematic uncertainty listed in Table I.
The uncertainty due to fitting for
$D_s^+\rightarrow\pi^+\pi^0$ is fractionally large 
because the signal yield is so small.}
\label{table:sys_BR}
\begin{tabular}{lcccccc}
\hline
Source  &  
$\frac{\displaystyle {\cal{B}}(K^+\pi^0)}{\displaystyle {\cal B}(\phi\pi^+)}$ &  
$\frac{\displaystyle {\cal{B}}(K^+\eta_{\gamma\gamma})}{\displaystyle {\cal B}(\phi\pi^+)}$ & 
$\frac{\displaystyle {\cal{B}}(K^+\eta_{3\pi})}{\displaystyle {\cal B}(\phi\pi^+)}$ & 
$\frac{\displaystyle {\cal{B}}(\pi^+\pi^0)}{\displaystyle {\cal B}(\phi\pi^+)}$ & 
$\frac{\displaystyle {\cal{B}}(\pi^+\eta_{\gamma\gamma})}{\displaystyle {\cal B}(\phi\pi^+)}$ &  
$\frac{\displaystyle {\cal{B}}(\pi^+\eta_{3\pi})}{\displaystyle {\cal B}(\phi\pi^+)}$  \\
\hline
Tracking  & 0.7  & 0.7 &  $-$ & 0.7 &  0.7 & $-$    \\
Particle identification  & 1.8 & 1.8 & 1.9 & 1.9 & 1.9 & 4.0  \\
$\pi^0$/$\eta \rightarrow \gamma\gamma$ & 2.4 & 2.4 & 2.4 & 2.4 & 2.4 & 2.4  \\
$O^{}_{\rm NN}$ requirement & 1.1 & 1.3 & 1.2 & 1.3 & 1.3 & 1.3  \\
$D_s^{*+}$ fraction in $\varepsilon$ & 0.7 & 0.7 & 0.7 &  0.7 & 0.7 & 0.7 \\
MC statistics & 0.8 &  0.8  & 0.8 & 0.8 & 0.7 &  0.7  \\
Fitting & 2.2 & 2.6 & 2.4 & 56.2 & 1.5 & 1.2 \\
$\mathcal{B} (\eta \rightarrow \gamma\gamma) $ & $-$ & 0.5 & $-$ & $-$ & 0.5 & $-$   \\
$\mathcal{B} (\eta  \rightarrow \pi^+ \pi^- \pi^0) $ & $-$ & $-$ & 1.2 & $-$ & $-$ & 1.2  \\
\hline
Overall uncertainty & 4.1 & 4.4 & 4.4 &  56.3 & 3.9 &  5.2  \\
\hline
\end{tabular}
\end{center}
\end{table*}

The systematic uncertainties for $A_{CP}$ are evaluated in
a similar manner as those for the branching fraction and
are summarized in Table~\ref{table:sys_CPV}.
The effect of a possible $CP$ asymmetry~\cite{PDG2020} 
in peaking background from 
$D^+\rightarrow \pi^+ (\pi^0/\eta)$ is 
considered as a systematic uncertainty.
The uncertainty in $A_{CP}$ due to 
our choice of $\cos\theta^{{\rm{CM}}}_{D_s}$ bins is 
evaluated by shifting the bin boundaries; the change in 
$A_{CP}$ is taken as the systematic uncertainty.
The uncertainty on $A_{CP}$ for the reference mode 
(taken from Ref.~\cite{PDG2020})
is taken as a systematic uncertainty.

\begin{table*}
\renewcommand{\arraystretch}{1.2}
\begin{center}
\caption{Systematic uncertainties for $A_{CP}$.
The overall uncertainty is the sum in quadrature of the listed uncertainties.}
\label{table:sys_CPV}
\begin{tabular}{lccccccc}
\hline
Source & & $K^+ \pi^0$  &   $K^+ \eta_{\gamma\gamma}$& $K^+ \eta_{3\pi}$ &  $\pi^+ \eta_{\gamma\gamma}$ &  $\pi^+ \eta_{3\pi}$ & $\phi \pi^+ $\\
\hline
Fitting  & & 0.0056 & 0.0035 & 0.0020 &  0.0005 & 0.0005 & 0.0002   \\
$D^+\rightarrow \pi^+ (\pi^0/\eta)$ background & & 
0.0062 & 0.0022 & 0.0031 &  $-$ & $-$ & $-$ \\
$\cos\theta^{\rm CM}_{D_s}$ binning & & 
0.0068 & 0.0028 & 0.0068 &  $-$ & $-$ & $-$ \\
$A_{CP}$ in $D_s^+ \rightarrow \phi \pi^+$ &  & $-$ & $-$ & $-$ &  0.0027 & 0.0027 & $-$  \\
\hline
Overall uncertainty & & 0.0108 & 0.0050 & 0.0077 & 0.0027 & 0.0027 & 0.0002\\
\hline
\end{tabular}
\end{center}
\end{table*}

In summary, we have used
the full Belle data set of 
921~fb$^{-1}$ to measure the branching fractions for four decay
modes of the $D_s^+$, and $CP$ asymmetries for three decay modes. 
Our results for the branching fractions relative to that of the 
reference mode $D_s^+\rightarrow\phi(\rightarrow K^+ K^-)\pi^+$ (${\cal B}^{}_{\phi\pi^+}$) 
are
\begin{eqnarray*}
{\cal B}(D_s^{+} \rightarrow K^{+}  \pi^0)/{\cal B}^{}_{\phi\pi^+} & = & 
(3.28 \pm 0.23 \pm 0.13 )\%    \\
{\cal B}(D_s^{+} \rightarrow K^{+} \eta)/{\cal B}^{}_{\phi\pi^+} & = & 
( 7.81 \pm 0.22 \pm 0.24 )\%     \\
{\cal B}(D_s^{+} \rightarrow \pi^{+} \pi^0) /{\cal B}^{}_{\phi\pi^+}  & = &  (0.16 \pm 0.25 \pm 0.09)\%   \\
{\cal B}(D_s^{+} \rightarrow \pi^{+} \eta)/{\cal B}^{}_{\phi\pi^+} & = & 
( 84.80\pm 0.47 \pm 2.64)\%  \,.
\end{eqnarray*}
Multiplying these results by the world-average value
${\cal B}^{}_{\phi\pi^+} = (2.24\pm 0.08)$\%~\cite{PDG2020} gives 
\begin{eqnarray*}
{\cal B}(D_s^{+} \rightarrow K^{+}  \pi^0) & = &  (0.735 \pm 0.052 \pm 0.030 \pm 0.026) \times 10^{-3}  \hspace*{0.50in}  \\
{\cal B}(D_s^{+} \rightarrow K^{+} \eta) & = &  (1.75 \pm 0.05 \pm 0.05 \pm 0.06)  \times 10^{-3}    \\
{\cal B}(D_s^{+} \rightarrow \pi^{+}\pi^0) & = & (0.037 \pm 0.055 \pm 0.021 \pm 0.001) \times 10^{-3}      \\
{\cal B}(D_s^{+} \rightarrow \pi^{+} \eta) & = &  (19.00 \pm 0.10 \pm 0.59 \pm 0.68) \times 10^{-3}  \,,
\end{eqnarray*}
where the third uncertainty listed is 
due to~${\cal B}^{}_{\phi\pi^+}$.
Our results for $D_s^{+} \rightarrow K^{+} \eta$ and 
$D_s^{+} \rightarrow \pi^{+}\pi^0$ are the most precise to date.
Our result for $D_s^{+} \rightarrow \pi^{+} \eta$ 
is consistent with a previous, less precise 
Belle result~\cite{Zupanc:2013byn}
and independent of it.
As we do not observe any signal for $D_s^+\rightarrow\pi^+\pi^0$, 
we set an upper limit on its branching fraction: 
\begin{eqnarray*}
{\cal B}(D_s^{+} \rightarrow \pi^{+}  \pi^0) & <  &  
1.2 \times 10^{-4} \hskip0.20in (90\%\ {\rm {C.L.}})\,.
\end{eqnarray*}
This is the most stringent constraint to date.

Our results for the $CP$ asymmetries are
\begin{eqnarray*}
A_{CP}(D_s^{+} \rightarrow K^{+}  \pi^0) & = &  0.064 \pm 0.044 \pm 0.011    \\
A_{CP}(D_s^{+} \rightarrow K^{+} \eta) & = &  0.021 \pm 0.021 \pm 0.004      \\
A_{CP}(D_s^{+} \rightarrow \pi^{+} \eta) & = &  0.002 \pm 0.003 \pm 0.003   \,.
\end{eqnarray*}
These results are also the most precise to date
and show no evidence of $CP$ violation.

\begin{acknowledgments}
We thank the KEKB group for the excellent operation of the
accelerator; the KEK cryogenics group for the efficient
operation of the solenoid; and the KEK computer group, and the Pacific Northwest National
Laboratory (PNNL) Environmental Molecular Sciences Laboratory (EMSL)
computing group for strong computing support; and the National
Institute of Informatics, and Science Information NETwork 5 (SINET5) for
valuable network support.  We acknowledge support from
the Ministry of Education, Culture, Sports, Science, and
Technology (MEXT) of Japan, the Japan Society for the 
Promotion of Science (JSPS), and the Tau-Lepton Physics 
Research Center of Nagoya University; 
the Australian Research Council including grants
DP180102629, 
DP170102389, 
DP170102204, 
DP150103061, 
FT130100303; 
Austrian Science Fund (FWF);
the National Natural Science Foundation of China under Contracts
No.~11435013,  
No.~11475187,  
No.~11521505,  
No.~11575017,  
No.~11675166,  
No.~11705209;  
Key Research Program of Frontier Sciences, Chinese Academy of Sciences (CAS), Grant No.~QYZDJ-SSW-SLH011; 
the  CAS Center for Excellence in Particle Physics (CCEPP); 
the Shanghai Pujiang Program under Grant No.~18PJ1401000;  
the Ministry of Education, Youth and Sports of the Czech
Republic under Contract No.~LTT17020;
the Carl Zeiss Foundation, the Deutsche Forschungsgemeinschaft, the
Excellence Cluster Universe, and the VolkswagenStiftung;
the Department of Science and Technology of India; 
the Istituto Nazionale di Fisica Nucleare of Italy; 
National Research Foundation (NRF) of Korea Grant
Nos.~2016R1\-D1A1B\-01010135, 2016R1\-D1A1B\-02012900, 2018R1\-A2B\-3003643,
2018R1\-A6A1A\-06024970, 2018R1\-D1A1B\-07047294, 2019K1\-A3A7A\-09033840,
2019R1\-I1A3A\-01058933;
Radiation Science Research Institute, Foreign Large-size Research Facility Application Supporting project, the Global Science Experimental Data Hub Center of the Korea Institute of Science and Technology Information and KREONET/GLORIAD;
the Polish Ministry of Science and Higher Education and 
the National Science Center;
the Ministry of Science and Higher Education of the Russian Federation, Agreement 14.W03.31.0026; 
University of Tabuk research grants
S-1440-0321, S-0256-1438, and S-0280-1439 (Saudi Arabia);
the Slovenian Research Agency;
Ikerbasque, Basque Foundation for Science, Spain;
the Swiss National Science Foundation; 
the Ministry of Education and the Ministry of Science and Technology of Taiwan;
and the United States Department of Energy and the National Science Foundation.
\end{acknowledgments}
\clearpage
\bibliography{ds_draft.bib}
\end{document}